\documentclass[%
floats,
 amsmath,amssymb,
 aps,
prb,
twocolumn,
floatfix,
]{revtex4-1}
\usepackage[pdftex]{graphicx}
\usepackage{dcolumn}
\usepackage{bm}
\usepackage{color}
\usepackage{hyperref}

\newcommand{\bq}{{\bf q}}
\newcommand{\bk}{{\bf k}}
\newcommand{\br}{{\bf r}}

\begin{document}
\title{Temperature-Dependent Band Structure of SrTiO$_3$ Interfaces}
\date{\today}
\author{Amany\ Raslan} 
\author{Patrick\ Lafleur} 
\author{W. A. Atkinson} 
\email{billatkinson@trentu.ca}
\affiliation{Department of Physics and Astronomy, Trent
  University, Peterborough Ontario, Canada, K9J 7B8} 
\begin{abstract}
We build a theoretical model for the electronic properties of the two-dimensional (2D) electron gas that forms at the interface between insulating SrTiO$_3$ and a number of polar cap layers, including LaTiO$_3$, LaAlO$_3$, and GdTiO$_3$.  The model treats conduction electrons within a tight-binding approximation, and the dielectric polarization via a Landau-Devonshire free energy that incorporates strontium titanate's strongly nonlinear, nonlocal, and temperature-dependent dielectric response. The self-consistent band structure comprises a mix of  quantum 2D states that are tightly bound to the interface, and quasi-three-dimensional (3D) states that extend hundreds of unit cells into the SrTiO$_3$ substrate.  We find that there is a substantial shift of electrons away from the interface into the 3D tails as temperature is lowered from 300~K to 10~K.  This shift is least important at high electron densities ($\sim 10^{14}$ cm$^{-2}$), but becomes substantial at low densities; for example, the total electron density within 4~nm of the interface changes by a factor of two for 2D electron densities $\sim 10^{13}$ cm$^{-2}$.  We speculate that the quasi-3D tails form the low-density high-mobility component of the interfacial electron gas that is widely inferred from magnetoresistance measurements.
\end{abstract}
\maketitle
\section{Introduction}
\label{sec:introduction}

At present, there is widespread interest in interfaces and heterostructures between SrTiO$_3$ (STO)  and polar perovksite materials such as LaAlO$_3$ (LAO).  Transition metal oxides are characterized by strong local interactions that often lead to novel magnetic, superconducting, or orbital-ordered phases that may be tailored by interface engineering.\cite{Zubko:2011ho}  The specific interest in STO was sparked by the observation of a two-dimensional electron gas (2DEG) at a LaTiO$_3$/STO interface,\cite{Ohtomo:2004hm} and by subsequent observations\cite{Reyren:2007gv,Brinkman:2007fk,Li:2011jx,Dikin:2011gl,Bert:2011,Kalisky:2012wf} that these 2DEGs exhibit ferromagnetism and superconductivity.  The ability to tune LAO/STO interfaces through metal-insulator and superconductor-insulator transitions by application of a gate voltage\cite{Thiel:2006eo,Caviglia:2008uh,Bell:2009eo,Liao:2011bk} has raised questions about the role of quantum criticality\cite{Schneider:2009gt} and the origins of superconductivity at low electron density.\cite{Edge:2015fj,Gorkov:2016dd,Ruhman:2016} 

 The 2DEGs reside primarily in the STO and extend very little into the cap layer,\cite{Popovic:2008ft,Son:2009wb,Delugas:2011ih} and consequently the basic elements of the electronic structure are similar for a variety of cap layer materials,\cite{Pentcheva:2007fn,Banerjee:2015wr,Chang:2013iq,Cancellieri:2013wa} and even for bare STO surfaces.\cite{SantanderSyro:2011hf,Meevasana:2011bh,Walker:2015}
Band structure calculations for LAO/STO interfaces\cite{Popovic:2008ft,Son:2009wb,Delugas:2011ih} predict that the majority of the conducting electrons reside in the TiO$_2$ planes adjacent to the interface, and occupy bands with $d_{xy}$ symmetry, while occupied bands with $d_{xz}$ and $d_{yz}$ symmetry extend farther into STO.   Because of the differences in their spatial extent, the $d_{xy}$ bands at the interface should be much more strongly affected by interfacial roughening than the $d_{xz}/d_{yz}$ bands,\cite{Popovic:2008ft}  and indeed Hall measurements have been interpreted in terms of a two-component system with two distinct mobilities.\cite{Kim:2010fl,Lerer:2011bp,Jost:2014uz,Joshua:2012bl,Guduru:2013iz}

A key feature of STO interfaces is that STO has an extremely high dielectric permittivity ($\epsilon \sim 10^4 \epsilon_0$, with $\epsilon_0$ the permittivity of free space) at low temperatures and weak electric fields, which strongly influences the profile of the charge density near the interface.  Importantly, $\epsilon$ is a strong function of temperature and electric field,\cite{Hemberger:1995dd,Dec:2005cr} so that the charge density profile can change dramatically with both temperature $T$ and  gate voltage.  To understand this, several calculations\cite{Copie:2009ev,Stengel:2011hy,Khalsa:2012fu,Park:2013gf,Gariglio:2015jx,Reich:2015ut,Peelaers:2015fh} have been made based on tight-binding or continuum models that build in relevant properties of the dielectric function.  These phenomenological approaches have tended to focus on  the nonlinear response of $\epsilon$ to the electric field as a way to understand the doping-dependence of the charge profile near the interface, and most ignore the nonlocal dielectric response that is inferred from the strong phonon dispersion at small wavevectors.\cite{Cowley:tr} One notable exception is Ref.~\onlinecite{Khalsa:2012fu}, which treats the lattice polarization ${\bf P}(\br)$ within a Landau-Devonshire approximation that inherently includes nonlocal and nonlinear effects. Ref.~\onlinecite{Khalsa:2012fu} focused on the doping dependence of the band structure at fixed temperature. Here, we extend their model to perform a systematic study of the temperature-dependent band structure of a generic STO interface. 

As with previous work,\cite{Copie:2009ev,Stengel:2011hy,Khalsa:2012fu,Park:2013gf,Gariglio:2015jx} we find that the 2DEG at the STO interface can be divided into a   quantum two-dimensional (2D) region that extends approximately 10 STO layers in from the interface, and a quasi-three dimensional (3D) region that extends deep into the STO.  The 2D quantum region is dominated by a band with $d_{xy}$ character that is weakly temperature-dependent at typical doping levels.  In contrast, the lowest energy $d_{xz/yz}$ bands extend much farther into the STO film, and are strongly temperature-dependent.  As a consequence, there is a substantial shift of charge away from the interface as temperature is lowered from 300~K to 10~K.   We show that this leads to  large differences in the photoemission spectra at low and high temperatures.  


The model employs a tight-binding approximation for the electrons, in which interactions are treated within a self-consistent field approximation.  The electrons couple to the polarization charge density $-\nabla\cdot {\bf P}$, where the polarization ${\bf P}$ is calcuated from a Landau-Devonshire energy that depends explicitly on temperature and electric field.  The model is agnostic about the doping mechanism, and simply assumes a confining potential at the interface due to a uniform external 2D charge density as one would expect from an electronic reconstruction,\cite{Nakagawa:2006gt} from oxygen vacancies at the LAO surface,\cite{Bristowe:2014fc} or from application of a gate voltage.  Alternative doping mechanisms such as O vacancies that accumulate at the interface\cite{Zhong:2010if,Bristowe:2014fc} are beyond the scope of this work.   Despite its complexity, this model neglects certain complicating aspects of the STO band structure that are not expected to change the broad trends identified in our calculations.  Notably, we ignore spin-orbit coupling, which mixes the different $t_{2g}$ orbitals and breaks the 3-fold $t_{2g}$ band degeneracy at the $\Gamma$ point.\cite{Caviglia:2010jv,BenShalom:2010kv,Khalsa:2012fu,Zhong:2013fv}  By so doing, we are able to study systems of up to 200 layers with a 2D grid of $200\times 200$ $\bk$-points; however, this means that some details of the band structure, especially at low charge densities, will be inaccurate.\cite{Khalsa:2012fu}  We have also ignored the renormalization of the band masses by electron-phonon\cite{Cancellieri:2016fw} and electron-electron interactions,\cite{Tolsma:2016dy} and the effects of  antiferrodistortive rotations of the unit cell below temperatures of 105 K.\cite{Mattheiss:1972dt,Tao:2016vv}  While these will affect our results quantitatively, the qualitative aspects of the calculations should be robust.

We describe the model in Sec.~\ref{sec:method}, and results of the calculations are given in Sec.~\ref{sec:results}.  First, the temperature-dependence of the charge distribution is described in Sec.~\ref{sec:temperature} for low, intermediate, and high electron densities (relative to typical experimental densities).  These results are then discussed in the context of the temperature- and doping-dependent band structure in Sec.~\ref{sec:bands}.  One direct experimental measure of the band structre is angle-resolved photoemission (ARPES), and in Sec.~\ref{sec:arpes} we focus on the implications of our calculations for ARPES.  We finish in Sec.~\ref{sec:local_v_nonlocal} with a brief examination of a particular aspect of our model, namely the role of nonlocal response in the dielectric function, which is shown to qualitatively affect the charge distribution near the interface at low temperatures.  Finally, in Sec.~\ref{sec:discussion} we propose that 3D tail states, which are ubiquitous in our calculations, form the high-mobility component of the electron gas that is widely observed in magnetotransport experiments.

\section{Method}
\label{sec:method}

\begin{figure}
    \centering
    \includegraphics[width=\columnwidth,natwidth=610,natheight=642]{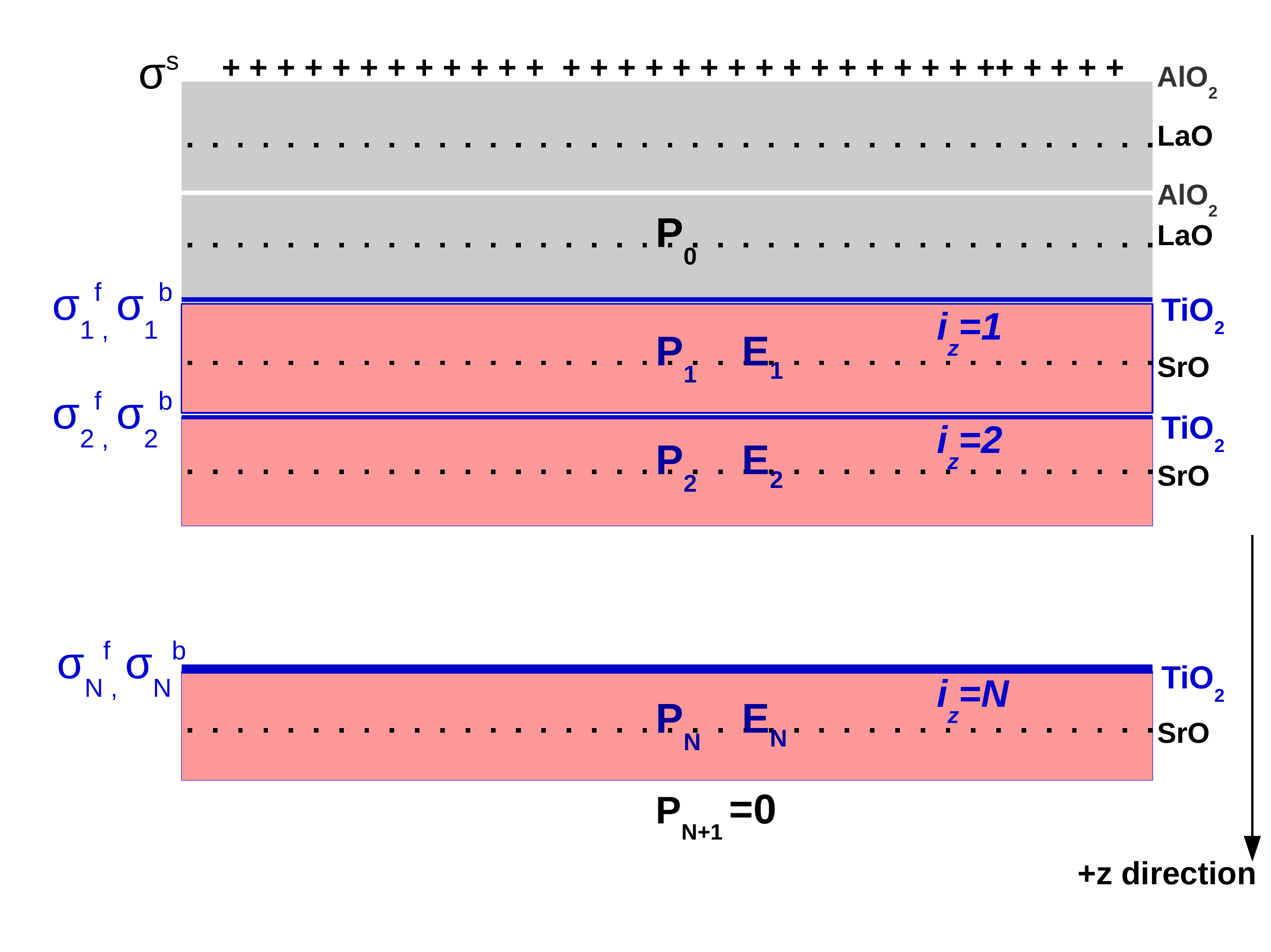}
        \caption{(Color online)
         Sketch of a model STO/LAO interface.  $N$ unit cells  of STO are stacked below an insulating LAO film in alternating TiO$_2$ and SrO  layers in the [001] direction.   Electronic reconstruction, gating, and surface O vacancies transfer charge from the top AlO$_2$ layer to the interface, leaving a residual 2D charge density $\sigma^s$ on the AlO$_2$ surface that attracts STO  conduction electrons to the interface.  The model is discretized along the $z$ direction, and assumes that the conducting TiO$_2$ layers are separated by blocks of dielectric; the polarization ${P_{i_z}}$ and electric field ${E_{i_z}}$ are therefore defined in the regions between the TiO$_2$ layers.  The conduction electrons in layer $i_z$ have 2D charge density $\sigma^f_{i_z}$, while the bound charge density due to the polarization is $\sigma^b_{i_z} =  P_{i_z} - P_{i_z+1}$.  We assume translational invariance in the planes, so the polarization, field, and electron density depend only on the layer index $i_z$.   An extra fictitious dielectric layer ($i_z = N+1$) is added to facilitate handling  the boundary condition $P_{N+1} = 0$  at the bottom of the STO substrate.  }
   \label{sketch}
\end{figure}

Our  interface model has two distinct pieces:  a self-consistent tight-binding description of the electronic bands and a Landau-Devonshire description of the polarization.  We begin with an overview of the model before discussing the two pieces in detail.  Figure \ref{sketch} shows the model's structure.  We consider a thick film of $N$ STO layers stacked in the [001] direction beneath an insulating cap layer.  In the figure, the cap is taken to be LAO, but our model only requires that it has a sufficiently wide band gap that it can be ignored.  We assume that some combination of  O-vacancy formation on the surface AlO$_2$ layer, electronic reconstruction, and application of a gate voltage transfers electrons to the STO interface, leaving a residual positive charge $\sigma^s$ (indicated by ``+'' signs) on the AlO$_2$ surface.  This residual charge creates an electric field that confines  the STO conduction electrons to the interface.

The model is discretized along the $z$ direction (perpendicular to the interface).    We treat the STO as a set of conducting TiO$_2$ planes separated by layers of dielectric.  As shown in Fig.~\ref{sketch}, the polarization and electric field are defined in the dielectric layers, while the charge density is confined to the 2D TiO$_2$ planes.  While the discretization process clearly limits the usefulness of the model at sub-unit cell length scales, it does nonetheless capture longer wavelength physics.  

We assume that we have translational invariance in the planar directions, so that the polarization, electric field, and charge density depend only on the layer index $i_z$.  Then, by symmetry, the polarization and electric field vectors  ${\bf P}$ and ${\bf E}$ must point in the $z$ direction.  The 2D charge density in the $i_z$th TiO$_2$ plane has two contributions:  a free charge density $\sigma^f_{i_z} $ due to the conduction electrons and a bound charge density $\sigma^b_{i_z} = P_{i_z} - P_{i_z+1}$ due to the polarization gradients.

We require boundary conditions for both the electric field and the polarization.  In the layered geometry, and for a fixed $\sigma^s$, the electric field in the STO is independent of the dielectric permittivity of the cap layer.  For simplicity, then, we take the polarization to be zero above the interface (ie. $P_0 = 0$) and the electric field above the first STO layer is therefore (by Gauss' law) $E_0 = \sigma^s/\epsilon_0$.    At large $z$, we expect the electric field and the polarization to be screened by the free charge density: to handle this, the electric field in the $N$th STO layer is zero (ie. $E_N=0$), and we add a fictitious $(N+1)$th layer in which $\sigma^f_{N+1} = P_{N+1} = 0$.

\subsection{Electronic Hamiltonian}
\label{sec:ham}
STO has a 3.3 eV band gap between filled O${2p}$ orbitals and empty Ti t$_{2g}$ orbitals.  For  an electron-doped interface we therefore include only the t$_{2g}$ orbitals in our model.  We adopt a tight-binding Hamiltonian with  three orbitals per unit cell, having $d_{xy}$, $d_{xz}$, and $d_{yz}$ character.     We consider only hopping between nearest-neighbor Ti atoms, and neglect matrix elements between orbitals of different symmetry:  these vanish in a cubic lattice, and are assumed small provided the lattice distortions are small.  Spin-orbit coupling also mixes different orbital types near band degeneracies, but as mentioned above, we gain a strong computational advantage by ignoring this effect.  We assume we have translational invariance with periodic boundary conditions in the $x$ and $y$ directions, and apply open (hard-wall) boundary conditions in the $z$ direction.

With these assumptions,  we  write the effective  Hamiltonian for the STO conduction electrons   as 
\begin{equation}
\label{H}
\hat H^{\mathrm{eff}}=\hat H_0 +\hat V^{\mathrm{ext}}+\hat V^{\mathrm{SC}}[\sigma^f,\sigma^b],
\end{equation}
where $\hat H_0$ is the tight-binding Hamiltonian for the inter-orbital hopping, 
$\hat V^{\mathrm{ext}}$ is the external potential energy due to the  charge at the LAO surface, and  $\hat V^{\mathrm{SC}}$ represents the self-consistent electrostatic potential energy due to both the free charge density $\sigma^f_{i_z}$ and the bound charge density $\sigma^b_{i_z}$ at the TiO$_2$ planes.   

The tight-binding term is
\begin{equation}
\hat H_0=\sum_{i_z,j_z}\sum_{\bf {k}}\sum_{\alpha\beta \sigma} c^\dagger_{i_z{\bf k}\alpha\sigma}{t}_{i_z\alpha,j_z\beta}({\bf k}) c_{j_z{\bf k}\beta\sigma},
\end{equation}
where $c_{i_z{\bf k}\beta\sigma}$ is the annihilation operator for an electron with spin $\sigma$ in layer $i_z$ and orbital type $\beta$,  ${\bf k}= (k_x,k_y)$ is a 2D wavevector, and 
 ${t}_{i_z\alpha,j_z\beta}({\bf k})$ is an element of the the tight-binding matrix, 
\begin{equation}
	{\bf t}({\bf k}) =\left [ \begin{array}{cccccc}  
	{\bf E}({\bf k}) & {\bf T}  & \ldots \\
	{\bf T} & {\bf E}({\bf k})\\
	   & &\ddots\\
	& & & {\bf E}({\bf k}) & {\bf T}\\
	& & & {\bf T} & {\bf E}({\bf k})
	\end{array}\right ] ,
\end{equation}
where ${\bf E}(\bk)$ and ${\bf T}(\bk)$ are matrices in the orbital basis,
\begin{eqnarray}
	{\bf E}({\bf k}) &=&\left [ \begin{array}{cccccc}  
	\xi_{xy}({\bf k}) & 0  & 0  \\
	0& \xi_{xz}({\bf k})&0\\
	0&0 & \xi_{yz}({\bf k})
	\end{array}\right ] \\
{\bf T} &=&\left [\begin{array}{cccccc}  
 -t^\perp & 0  & 0 \\
	0& -t^\parallel&0\\
	   0&0 &-t^\parallel
	\end{array}\right],
\end{eqnarray}
and
\begin{eqnarray}
	\nonumber
	&&\xi_{xy}({\bf k})=\epsilon_{t_{2g}}-2t^\parallel (\cos k_x a+\cos k_y a ),\\
	&&\xi_{xz}({\bf k})=\epsilon_{t_{2g}}-2t^\parallel \cos k_x a-2t^\perp \cos k_y a, 
		\label{eq:ea}\\
	&&\xi_{yz}({\bf k})=\epsilon_{t_{2g}}-2t^\perp \cos k_x a-2t^\parallel \cos k_y a, \nonumber\end{eqnarray}
are planar dispersions.  Here, $\epsilon_{t_{2g}}$ is the on-site orbital energy (which can be set to 0), and $a$ is the STO  lattice constant.  For a given symmetry of $t_{2g}$ orbital there are two distinct hopping processes between nearest-neighbor Ti atoms:  the hopping amplitude is $t^\parallel$ between Ti atoms  in the same plane as the orbital (eg.\ the $x$-$y$ plane for $d_{xy}$ orbitals), while it is $t^\perp$ perpendicular to the plane of the orbitals (eg.\ along the $z$ direction for $d_{xy}$ orbitals).
Since nearest-neighbor $d_{xy}$ orbital wavefunctions overlap more in the $x$-$y$ plane 
than along the $z$-axis,  $t^\parallel  \gg t^\perp $.  Values for 
$t^\|$, $t^\perp$,  and other model parameters are given in Table~\ref{cons}. 

\begin{table} 
 \begin{tabular}{l | r}
  \hline
  \multicolumn{2}{c}{Model parameters} \\
  \hline
 $t^\parallel $& 0.236 eV\\
  $t^\perp$ & 0.035 eV\\
 $a$ & 3.9 \AA \\
 M & 24 amu\\
 Q & $8.33e$ \\
  $\omega_0$ & $2.5 \times 10^{13}$ s$^{-1}$ \\
  $\omega_1$ & $1.7 \times 10^{13}$ s$^{-1}$ \\
  $\alpha_1$ & $1.15a$ \\
 $\alpha_2$ & $5a$  \\
 $\epsilon_\infty$ & $5.5\epsilon_0$\\
$T_0$ & $1.46\times 10^4$ K\\
    ${T_s}$ & 15 K\\
    ${\xi}$ & 1.45\\ 
$\gamma$ & 63 eV$\cdot$\AA$^{-4}$ \\
  \hline
\end{tabular}
  \caption{Model parameters used in our calculations.  Values are taken from Ref.~\onlinecite{Khalsa:2012fu} except for $T_0$, ${\xi}$, ${T_s}$, and ${\gamma}$, which are obtained by fitting to the temperature- and field-dependence of the experimental dielectric susceptibility (Appendix \ref{app:A}).}
  \label{cons}
\end{table}

Assuming that  the LAO surface charge is uniformly distributed, we obtain a simple description for the potential energy of an electron in the confining field,
\begin{equation}
\hat V^{\mathrm{ext}}= \frac{\sigma^s e }{2\epsilon_{\infty}}\sum_{\bf {k}}\sum_{i_z\alpha\sigma} z c^\dag_{i_z\bf {k}\alpha\sigma}c_{i_z\bf {k}\alpha\sigma},
\end{equation}
where $\epsilon_\infty$ is the optical dielectric constant due to electronic screening, and $z= i_z a$ is the distance from layer $i_z$ to the interface. 

Under a similar assumption, the self-consistent potential energy due to both the 2DEG  and the 2D bound charge density is
\begin{eqnarray}
\nonumber
\hat V^{\mathrm{SC}}[\sigma^f,\sigma^b]&=& \frac{e}{2 \epsilon_\infty } \sum_{\bf {k}}\sum_{i_z\alpha\sigma}\sum_{j_z} [\sigma^f_{j_z}+\sigma^b_{j_z}]\\
&&\times (\vert z^\prime -z \vert- z)c^\dag_{i_z\bf {k}\alpha\sigma}c_{i_z\bf {k}\alpha\sigma}, 
\label{eq:Vsc}
\end{eqnarray}
where   $z^\prime =j_z a$,  $\sigma^f_{j_z}=  -e \sum_\beta n_{j_z\beta}/a^2$ is the 2D charge density in layer $j_z$ and   $n_{j_z\beta}$ is the electron occupation number for orbitals of type $\beta$ in layer $j_z$.  The charge density is calculated self-consistently from
\begin{equation}\label{n}
n_{j_z\beta}=\frac{2}{N_{\bf k}} \sum_{{\bf k}} \sum_{n} | \psi_{j_z \beta,n}({\bf k})|^2 f(\epsilon_{n{\bf k}}),
\end{equation}
where the factor of 2 is for spin, $\epsilon_{n\bk}$ and $\psi_{j_z \beta,n}({\bf k}) $ are the energy eigenvalues and eigenstates of $\hat H^\mathrm{eff}$ respectively, and $ f(\epsilon_{n{\bf k}})$ is the Fermi-Dirac distribution function.  The bound charge density is $\sigma^b_{j_z} = P_{j_z} - P_{j_z+1}$, where the polarization $P_{j_z}$ is obtained from the Landau-Devonshire model discussed in the next section.

Because we have neglected contributions to the Hamiltonian that mix different orbital symmetries, each band has a well-defined orbital character. As a consequence, the band index $n$ can be written in the form $\tilde n\alpha$ where $\alpha$ is one of $xy$, $xz$, or $yz$ and $\tilde n$ is an integer labeling  bands of type $\alpha$ (the $1xy$ band is the lowest-energy $xy$ orbital character band, etc.).  Furthermore, the lack of orbital mixing leads to a particularly simple form of the Hamiltonian such that the eigenvectors $\psi_{j_z \beta,n}(\bk)$ are independent of $\bk$, and  the eigenvalues obtain the simple form
\begin{equation}
\epsilon_{\tilde n\alpha \bk} = \epsilon_{\tilde n\alpha \bk=0}+ \xi_\alpha(\bk) - \xi_\alpha(0),
\end{equation}
where $\xi_\alpha(\bk)$  is given by Eq.~(\ref{eq:ea}). It is thus possible to determine the spectrum at finite $\bk$ from the energy eigenvalues at $\bk=0$, and we therefore only need to diagonalize the Hamiltonian once per $\bk$-sum.  The resulting speed-up allows us to study large system sizes of up to 200 layers with $200 \times 200$ $k$-points.

\subsection{Polarization Model}
\label{sec:polarization}
The high polarizability of STO is due to the presence of a soft transverse optical phonon mode that is associated with an incipient ferroelectric transition.  The transition is suppressed by quantum fluctuations, so that the dielectric susceptibility saturates at a characteristic temperature $T_s \sim 15$ K.  Here, the induced polarization $P_i$ is defined for unit cell $i = (i_x,i_y,i_z)$ in terms of the normal-mode coordinate $u_i$ and effective charge $Q$ associated with the soft mode via
\begin{equation} 
 P_i=\frac{Qu_i}{a^3}.
 \label{eq:Pu}
 \end{equation}
 The normal-mode coordinate represents the amplitude of the lattice distortion, projected onto the soft optical phonon eigenvector,\cite{Lines:2001bn} and $Q$ is a fitting parameter that relates the collective coordinate to the polarization (see Table~\ref{cons}).  As discussed above, translational symmetry in the $x$-$y$ plane ensures that $u_i$ and $P_i$ are polarized along the $z$ direction, and that they are independent of $i_x$ and $i_y$. 

The polarization is obtained by minimizing a simple free energy that includes temperature, electric field, and nonlocal effects.  Model parameters have been set by fitting to temperature- and field-dependent dielectric susceptibility measurements of Ref.~\onlinecite{Dec:1998} while the nonlocal correlations are inferred from neutron scattering measurements of the phonon dispersion.\cite{Cowley:tr} The fitting process is discussed in Appendix~\ref{app:A}, and
the model reproduces the measured differential susceptibility with a maximum relative error of $16\% $ for temperatures $0 \leq T \leq 70$ K and $0 \leq E \leq 500$ V/mm; the relative error is $6\% $ at room temperature.

The simplest quartic free energy has the form\cite{Khalsa:2012fu}
\begin{equation}\label{pot}
\frac{U}{N_{2D}}=\frac{1}{2}\sum_{i_z,j_z} u_{i_z}{D}_{i_zj_z}{u}_{j_z}-Q\sum_{{i_z}}{E}_{i_z}{u}_{i_z}+\frac{\gamma}{4}\sum_{i_z}{u}_{i_z}^4,
\end{equation}
where $N_{2D}$ is the number of 2D unit cells in the $x$-$y$ plane, $D_{i_zj_z}$ is a matrix that contains the force constants between layers $i_z$ and $j_z$,  $E_{i_z}$ is the electric field, and ${\gamma}$ is a constant of proportionality for non-linear response.  This latter term is important only at high electron densities where the electric field is very strong. 

 The potential energy can be then minimized by taking the derivative with respect to $u_{l_z}$ and setting it equal to zero, from which we obtain the constituent equation
\begin{equation}
QE_{l_z}=\sum_{j_z}D_{l_z j_z }u_{j_z} +\gamma u_{l_z}^3.
\label{u}
\end{equation}
for $u_{l_z}$. Here, the electric field $E_{l_z}$ is equal to minus the gradient of the  total electric potential, which contains contributions from the external surface charge, the bound polarization charge, and the free electron charge.

To obtain $D_{l_zj_z}$, we Fourier transform the phenomenological expression proposed in Ref.~\onlinecite{Khalsa:2012fu},
\begin{equation}
D_{q_z}=M \left [\omega^2_0-\omega_1^2e^{-(\alpha_1 q_z)^2/2}-\omega_2^2(T)e^{-(\alpha_2 q_z)^2/2}\right ],
\label{eq:Dkz}
\end{equation} 
to model the dispersion  of the ferroelectric phonon mode, given by $\omega_{q_z}(T) = \left [ D_{q_z} /M\right]^{1/2}$.
The parameter values for $\omega_0$, $\omega_1$, and $\alpha_1$ (Table~\ref{cons}) are taken from Ref.~\onlinecite{Khalsa:2012fu}, 
but   $\omega_2(T)$ and $\alpha_2$ are modified to fit the low temperature dielectric susceptibility.

For $\omega_2(T)$, we take the phenomenological form (Appendix~\ref{app:A})
\begin{equation}
\omega_2^2(T) =\omega_0^2-\omega_1^2-\frac{Q^2 T_Q^\xi}{M \epsilon_0 a^3 T_0^\xi},
\label{eq:w2T}
\end{equation}
where $T_Q = T_s \coth(T_s/T)$ is an effective temperature that incorporates quantum fluctuations of the ferroelectric phonon mode.\cite{Kleemann:1998ut}  
The power $\xi = 1.45$  is chosen to improve the quantitative fit to experiments and the constant $T_0$ is obtained from the zero-field susceptibility $\chi(T) = (T_0/T_Q)^\xi$.   
While the effective temperature reduces to $T_Q = T$ at high temperatures, giving a Curie-like susceptibility, 
it saturates at $T_Q = T_s$ at low temperatures; consequently, the divergence at $T=0$ is avoided and the susceptibility  saturates at $\chi(T\rightarrow0) = (T_0/T_s)^\xi$.

In summary, the self-consistency cycle for $\sigma^f_{i_z}$ and $\sigma^b_{i_z}$ involves solving Eqs.~(\ref{n}) and (\ref{u}) for a given electric field to obtain the electron density and lattice polarization, and then updating the electric field from the resulting potential.  As has been pointed out elsewhere, the self-consistency cycle is numerically unstable\cite{Khalsa:2012fu}, and to address this we have implemented Anderson mixing of the electric potential.\cite{Eyert:1996gv} In addition, we have found that convergence is most readily obtained if the initial guess for the simulations takes the electron density to be $-\sigma^s$ in the 1st STO layer and zero elsewhere.

\section{Results}
\label{sec:results}


In this section,  we present the results of our calculations for temperature and doping-dependent electronic structure  of the LAO/STO interfaces. Early DFT calculations established\cite{Gariglio:2015jx}  that the interface breaks the cubic symmetry of the ideal STO lattice, so that a qualitative difference emerges between $d_{xy}$ orbitals (which are oriented parallel to the interface) and $d_{xz/yz}$ orbitals.  The hopping amplitude along the $z$ axis is $t^\perp$ for $d_{xy}$ orbitals and $t^\|$ for $d_{xz}$ and $d_{yz}$ orbitals.  Since $t^\| \sim 10 t^\perp$, this corresponds to an effective mass along the $z$ direction that is 10 times larger for $xy$ bands than for $xz$ or $yz$ bands.  This difference sets the energy ordering of the bands, such that the lowest-energy band has $xy$ symmetry and is tightly confined to within a few unit cells of the interface;  the lowest $d_{xz/yz}$ bands are higher in energy and extend farther from the interface. 

In an ideal polar catastrophe model, a charge transfer of $0.5$ electrons per unit cell is needed to suppress the potential divergence in the  polar cap material. The ideal value of  0.5$e/a^2$ has been measured for GdTiO$_3$/SrTiO$_3$ interfaces,\cite{Moetakef:2011ko} and only sporadically in LAO/STO interfaces;\cite{Guduru:2013iz,Jost:2014uz}  in  most conducting interfaces  typical experimental values of the electron density measured by the Hall effect \cite{Dubroka:2009,Cancellieri:2013wa} range from $10^{13}$ to $10^{14}$ $e$/cm$^{2}$.    The charge density can be further modulated by a gate voltage, and we therefore  perform calculations for three different doping levels that cover common  experimental and theoretical values of the $2$D charge density:   $\sigma^s = 0.5e/a^2$ ($3.3\times 10^{14}$ $e$/cm$^{2}$),  as predicted by the polar catastrophe model;  $\sigma^s = 0.1e/a^2$ ($6.5 \times 10^{13}$ $e$/cm$^{2}$), which is a typical doping found in LAO/STO interfaces; and  $\sigma^s =0.05e/a^2$ ($3.3 \times 10^{13}$ $e$/cm$^{2}$), which is approaching the metal insulator transition that is observed at $\sim 10^{13}$~$e$/cm$^{2}$.  Several calculations have explored the doping dependence of the electronic structure at low $T$,\cite{Copie:2009ev,Stengel:2011hy,Khalsa:2012fu,Park:2013gf,Gariglio:2015jx}
and we observe similar trends with doping in our low-T calculations. The main new results at this paper refer to how the T-dependence of the electronic structure evolves with doping.

\subsection{Effect of temperature on the charge distribution}
\label{sec:temperature}
 In this section, we examine the temperature-dependence of the charge distribution for the three representative cases listed above.  To minimize finite-size effects, all calculations are for an STO slab of thickness $L=200$ layers (see Appendix~\ref{sec:fs}).  We show that there is a pronounced shift of charge density from 2D quantum states that are confined to within $\sim 4$ nm of the interface into 3D tail states that extend hundreds of unit cells into the STO; the degree of this shift depends strongly on doping.

\begin{figure}[tb]
    \centering
    \includegraphics[width=\columnwidth,natwidth=610,natheight=642]{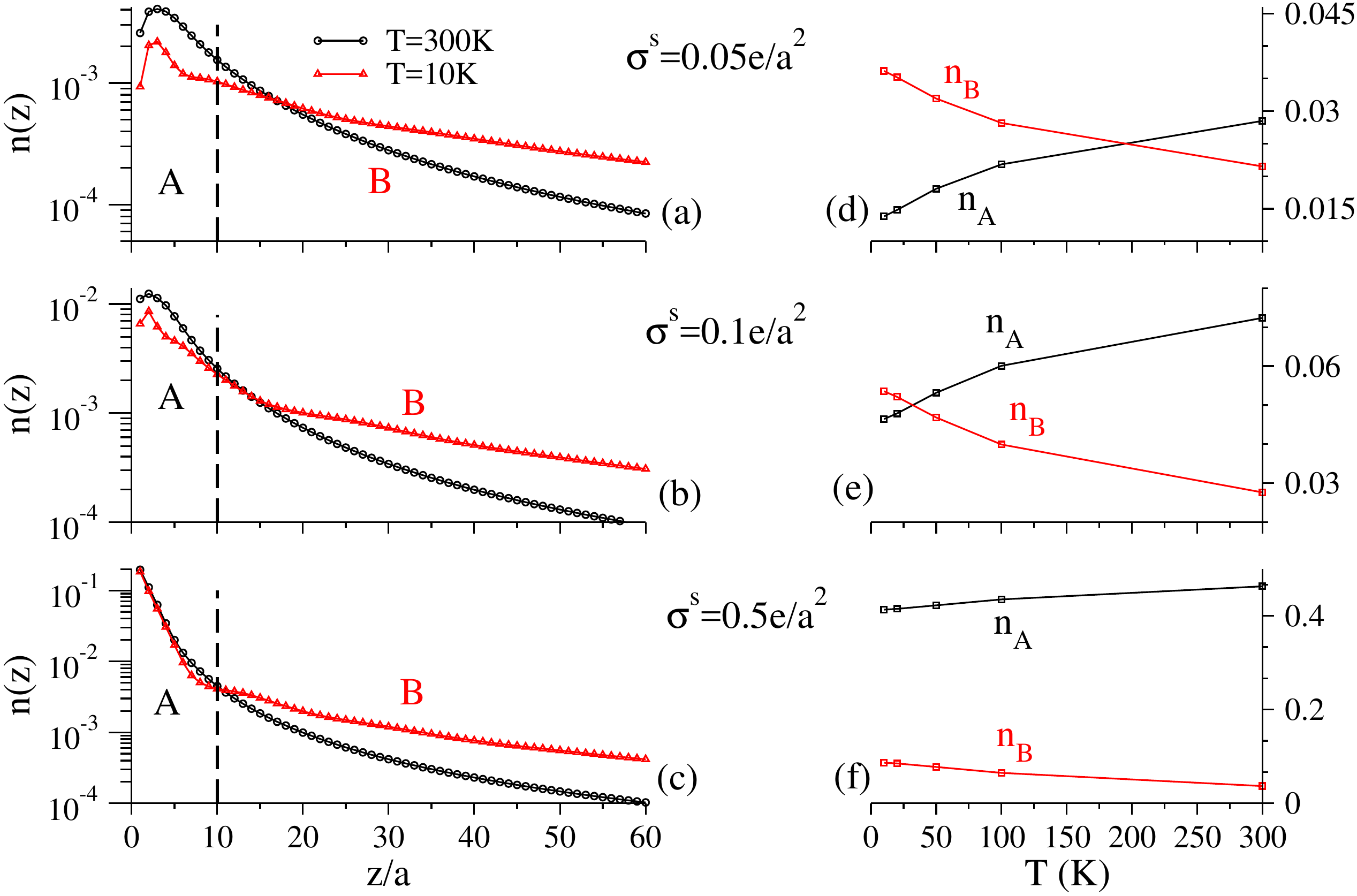}
    \caption{(Color online) Electron density $n(z)$ per unit cell inside an STO slab at different temperatures and dopings. Results are for (a) $\sigma^s= 0.05e/a^2$, (b) $0.1e/a^2$, and (c) $0.5 e/a^2$ at $T=10$~K and $T=300$~K.  The vertical dashed lines define regions A ($z \leq 10a$) and B ($z > 10a$), which roughly correspond to the interface and tail regions.    (d)-(f) The total 2D electron density in regions A and B as a function of temperature.   The figure shows the first 60 layers of an $L=200$ layer STO slab. }
   \label{n01}
\end{figure}

 Figure~\ref{n01}(a)-(c) shows the electron density, $n(z)=\sum_{\beta} n_{i_z\beta}$ (where $z=i_za$),  for 10~K and 300~K and for low ($0.05e/a^2$), intermediate ($0.1e/a^2$), and high ($0.5e/a^2$) electron densities.     As we discuss below, the charge distribution is a mix of surface states with strongly 2D character and tails with 3D character. This  is particularly evident in  the low-$T$ results in Fig.~\ref{n01}, which show a clear distinction between surface and  tail regions.  At high $T$, the distinction blurs, and $n(z)$ drops off rapidly in the tail region.  The crossover between surface and tail occurs at $z \approx 10a$ ($z\approx 4$ nm), and for discussion purposes we divide the profile into region A  ($z \leq 10a$) and region B  ($z >10a$). The charge densities $n_A$ and $n_B$ for each region are plotted as a function of $T$ in Fig.~\ref{n01}(d)-(f).

There are two key points made by Fig.~\ref{n01}.  The first  is that the fraction of the total electron density in region A depends on $\sigma^s$.    At 300~K,  about 90\% of the charge lies in region A  for high $\sigma^s$, whereas only about half of the total charge lies in region A at  low $\sigma^s$.  

The second point is that, except at the highest doping levels,  $n(z)$ depends strongly on $T$:  the charge density near the interface decreases as the temperature is lowered while it increases in the tails.      The contrast between low and high charge densities is striking:   $n_A$ doubles between 300~K and 10~K for low charge density ($\sigma^s = 0.05e/a^2$), but changes by only 10\% for high charge density  ($\sigma^s = 0.5e/a^2$).   Focusing on the middle ``typical'' value of $\sigma^s = 0.1e/a^2$, we note that   about  70\% of the total electron density lies in region A at 300 K, in agreement  with Ref.~\onlinecite{Copie:2009ev}, and slightly under half remains at 10~K.

One of the most striking features of Fig.~\ref{n01} is that the profile of $n(z)$ near the interface is almost independent of $T$ at the highest charge density, but is strongly $T$-dependent at the lowest charge density.  This trend is connected to the nonlinearity of the dielectric response in strong electric fields.  When $\sigma^s$ is large, the electric fields near the interface are large, and the nonlinear term ($\gamma u_{l_z}^3$) in Eq.~(\ref{u})  dominates the linear term ($\sum_{j_z}D_{l_z j_z }u_{j_z} $).  Because we have taken $\gamma$ to be $T$-independent, $n(z)$ is also $T$-independent in this region.  The electric field decreases both as one moves away from the interface, and as one decreases $\sigma^s$; in both regimes, $n(z)$ becomes temperature-dependent because the nonlinear contribution to the  dielectric response is small. 

It should be noted that in the nonlinear regime, the lattice polarization due to an electric field is proportional to  $\gamma^{-1/3}$ [from Eq.~(\ref{u})], so that $\gamma$ must change by a relatively large amount to have a significant effect on the charge distribution.  Indeed, $\gamma$ has been measured experimentally\cite{Dec:2005cr} below 60~K  and was found to be roughly constant down to 30~K, and then to increase by about 50\% as the system was further cooled.  This corresponds to a change of only 15\% in the nonlinear dielectric screening.  Unless $\gamma$ changes significantly at higher $T$, the assumption of constant $\gamma$ is  reasonable.    
 
 \begin{figure}[tb]
    \includegraphics[width=\columnwidth]{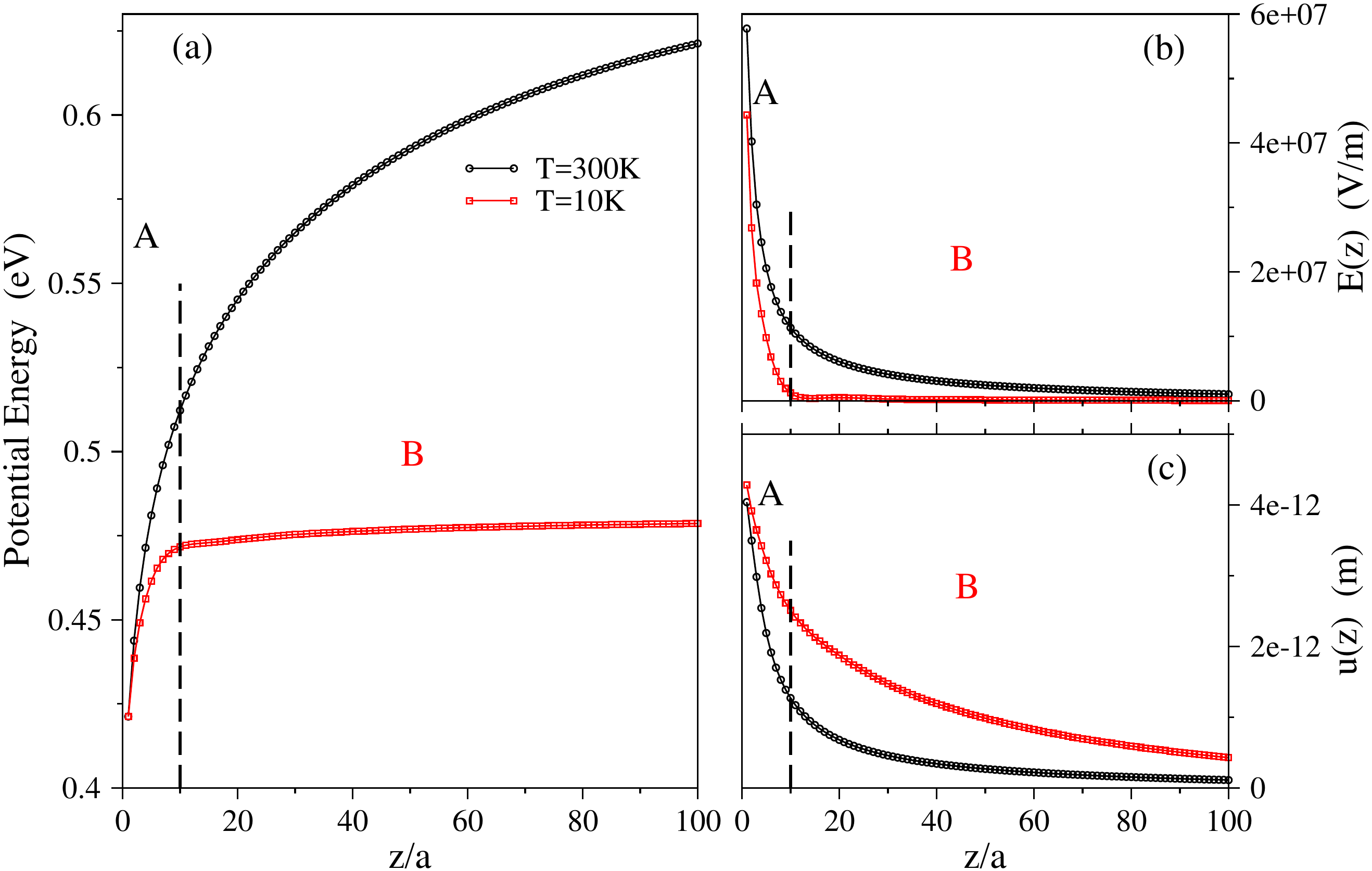}
    \caption{(Color online) Details of the self-consistent solution at low and high temperature for $\sigma^s = 0.1e/a^2$.  (a) The self-consistent potential energy, (b) the electric field, and (c) lattice normal mode displacement are shown at 300 K and 10 K.}
   \label{V01}
\end{figure}

 To understand better the charge deconfinement that occurs at low temperatures, we plot the electronic potential energy (ie.\ the electron charge times the potential), the electric field, and the normal mode displacement at high and low temperatures in Fig.~\ref{V01} for the intermediate value of $\sigma^s$.    Figure~\ref{V01}(a) shows that, in region A, there is a triangular quantum well that confines electrons in 2D quantum states near the interface at all temperatures.  In contrast, the potential in region B is strongly  temperature dependent, with a crossover from a deep well at high temperature to a nearly flat potential at $10$K.   This strong $T$-dependence is connected to the linear dielectric function, which changes by two orders of magnitude between 300~K ($\epsilon \approx 300 \epsilon_0$)  and 10~K ($\epsilon \sim 10^4\epsilon_0$). 

Because of the large value of $\epsilon$,  the electric field is strongly screened in region B at low temperature [Fig.~\ref{V01}(b)].   According to Gauss' law, 
\begin{equation}
\epsilon_\infty \frac{\partial E(z)}{\partial z} =-en(z)-\frac{\partial P(z)}{\partial z}
\end{equation}
where $P(z)$ is the lattice polarization, $E(z)$ is the electric field, and $\epsilon_\infty= 5.5\epsilon_0$ the optical dielectric constant.
Because the electric field  is small in region B,
we have 
\begin{equation}
en(z)\approx -\frac{\partial P}{\partial z},
\end{equation} 
at $T=10$~K.
  This means that the electric field generated by the conduction electrons in region B is nearly compensated by the lattice polarization.  The normal coordinate $u(z)$ for the soft phonon mode, which is related to the polarization by Eq.~(\ref{eq:Pu}),  is shown in Fig.~\ref{V01}(c).  Here, we see that $u(z)$ decays with $z$ more slowly at low $T$ than it does at high $T$, consistent with enhanced dielectric screening at low $T$.

\begin{figure}[tb]
    \centering
    \includegraphics[width=\columnwidth,natwidth=610,natheight=642]{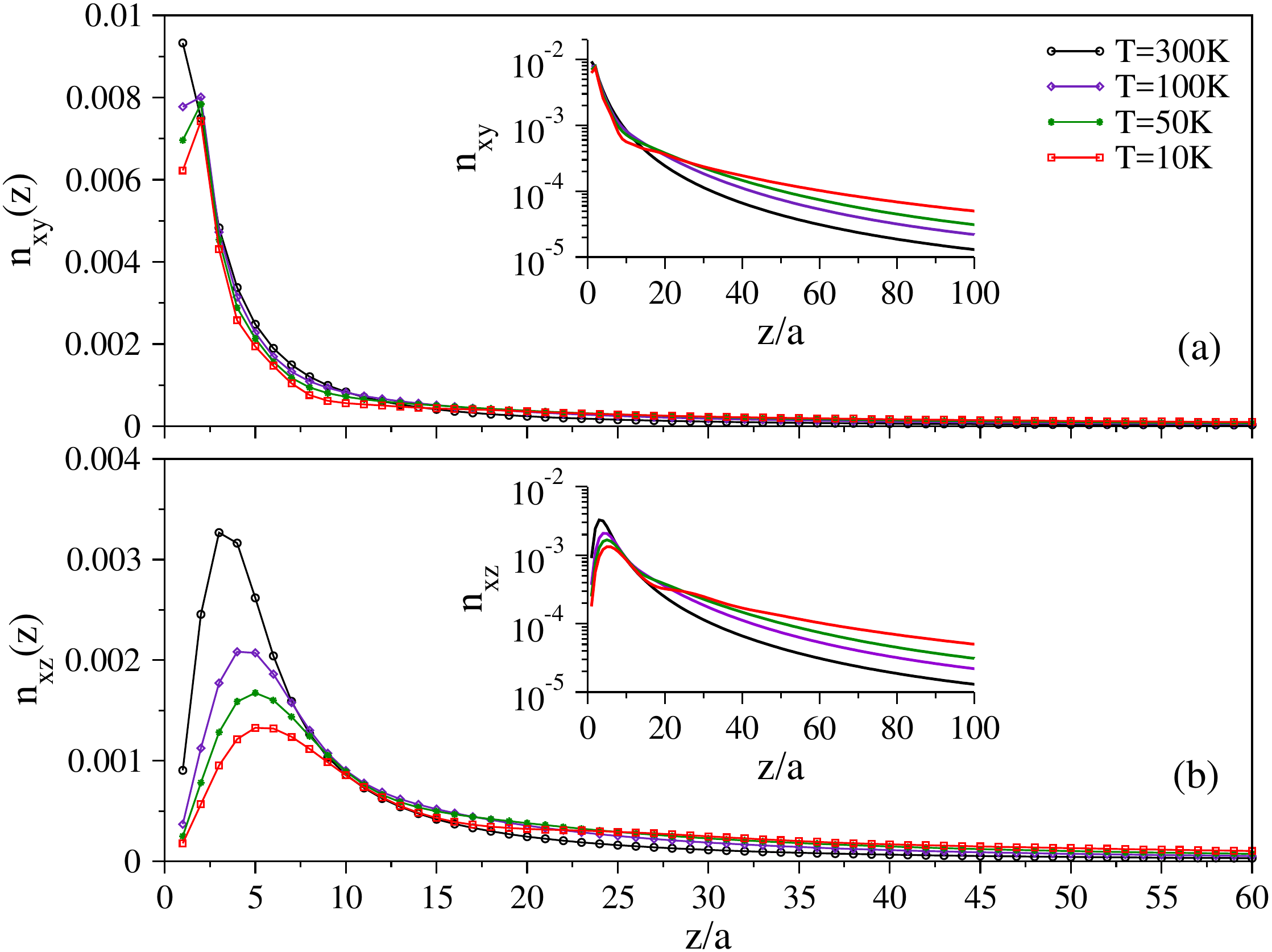}
    \caption{(Color online)  Electron density per unit cell $n_\alpha(z)$ for orbital types (a) $\alpha = xy$, and (b) $\alpha=xz$.  Results are at temperatures range from 300~K to 10~K.  Note that $n_{yz}(z) = n_{xz}(z)$.   Insets show the electron density on  logarithmic scale.  Results are for $\sigma^s = 0.1 e/a^2$.}
   \label{nxyz}
\end{figure}

For completeness, we plot the charge density for intermediate doping as a function of orbital type in  Fig.~\ref{nxyz}.  This figure shows that, while the interfacial $d_{xy}$ electron density $n_{xy}(z)$ is weakly temperature dependent, $n_{xz}(z)$ and $n_{yz}(z)$ evolve strongly with $T$ near the interface.  In particular, the $d_{xz}$ and $d_{yz}$ bands combined account for 80\% of the charge  transfer out of the first 10 layers as the temperature decreases.  The different sensitivities of $n_{xy}(z)$ and $n_{xz/yz}(z)$ to temperature follow from the different mass anisotropies of the three bands:  both the $xz$ and $yz$ bands are light along the $z$ direction while the $xy$ bands are heavy;  the $xz$ and $yz$ wavefunctions are therefore more extended along $z$ than the $xy$ wavefunctions.  It is  unsurprising that the $xz$ and $yz$ bands are most affected as the confining potential weakens when $T$ is reduced.

In summary, we arrive at the following scenario:  at room temperature, a  majority of electrons is confined to quantum states within $\sim 4$~nm of the interface by strong electric fields associated with the surface charge $\sigma^s$;  however, as $T$ is reduced, this electric field is increasingly screened by the dielectric response of the STO, causing a partial deconfinement of the electron gas. This deconfinement is most pronounced at the lowest $\sigma^s$, where approximately half of the interfacial electron density moves into the tail region.  Despite the large fraction of electrons in the tails, the associated electric fields are vanishingly small because of the strong dielectric screening.  We note in passing that the structure of the tails, and in particular the connection to ferroelectric quantum criticality in the STO substrate, is discussed in detail in Ref.~\onlinecite{Atkinson:2016}

\subsection{Effect of temperature on the band structure}
\label{sec:bands}
\begin{figure}[tb]
 \center
  \includegraphics[width=\columnwidth,natwidth=610,natheight=642]{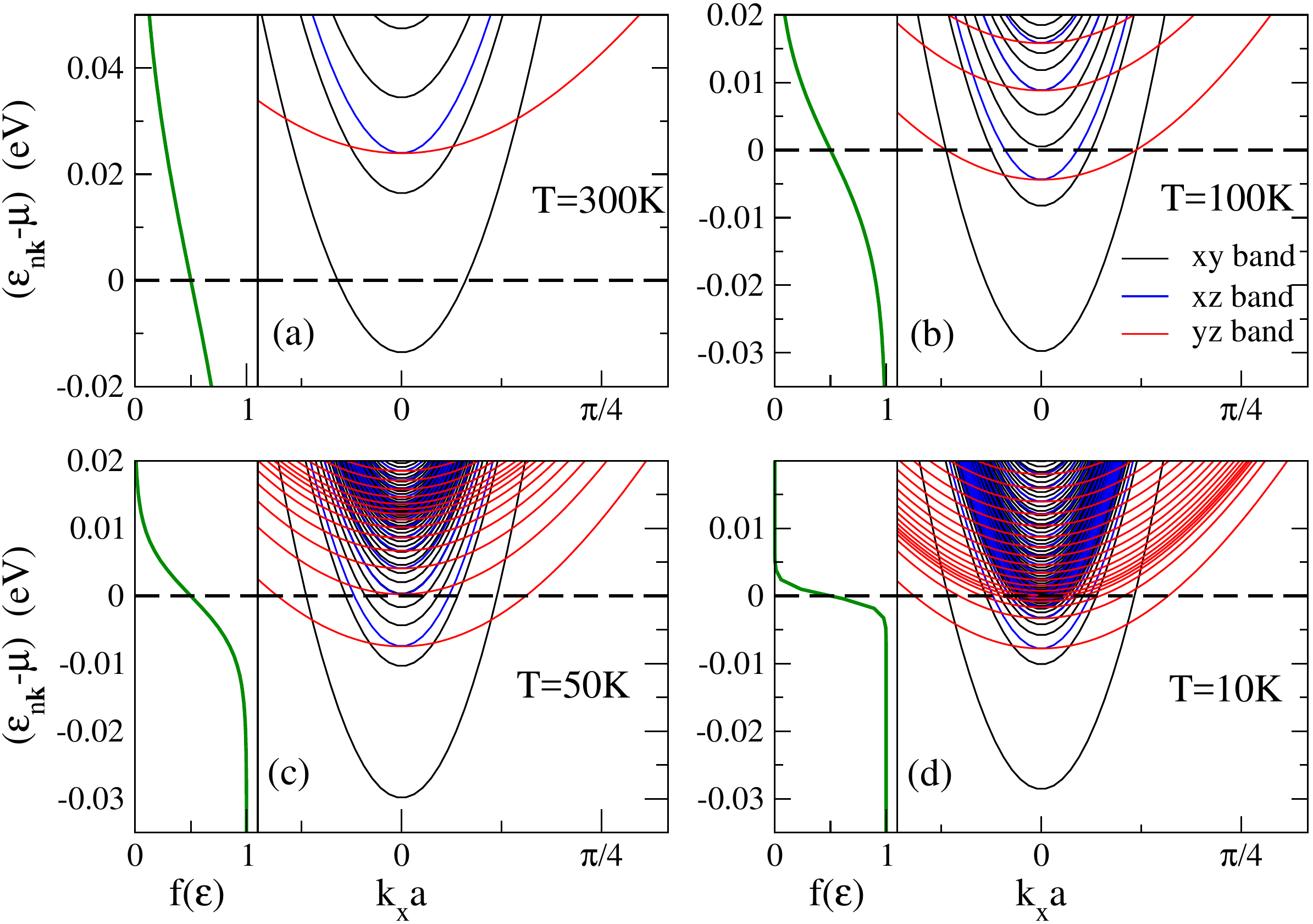}
    \caption{(Color online)  Self-consistent band structure along $\bk = (k_x,0)$. Results are for (a) 300~K, (b) 100~K, (c) 50~K, and  (d) 10~K, and $\sigma^s = 0.1e/a^2$. The Fermi-Dirac distribution function, $f(\epsilon)$, is shown in each panel (green line).}
   \label{band01}
\end{figure}

The temperature-dependent band dispersions $\epsilon_{n\bk}$ are shown in  Fig.~\ref{band01} for intermediate charge density.  The t$_{2g}$ orbital degeneracy is broken by the interface, resulting in multiple orbitally polarized sub-bands.\cite{Popovic:2008ft} The sub-bands consist of light bands (black lines) with $d_{xy}$ orbital character, and two anisotropic bands (blue and red lines) with $d_{xz}$ and $d_{yz}$ orbital character.     At all temperatures, the two lowest-energy sub-bands at $\bk = 0$ have $d_{xy}$ orbital character, while $d_{xz}$ and $d_{yz}$ sub-bands appear at higher energies.  This structure is consistent with previous  DFT calculations \cite{Stengel:2011hy,Zhong:2013fv} and with photoemission experiments.\cite{Walker:2015}

Figure~\ref{band01}(a) shows the $1xy$, $2xy$,  $1xz$, and $1yz$ sub-bands at 300 K.   We note that while the electrochemical potential $\mu$  lies below all but the  $1xy$  band at 300~K, the thermal energy is sufficient that all  bands shown in Fig.~\ref{band01}(a) have significant electron occupation.  The $1xy$ band has the highest occupancy, containing about $20\%$ of the total electron density,  while the first four bands combined contain approximately half of the total charge.

Two significant changes occur as the temperature is lowered: first, there is a significant shift of the electrochemical potential $\mu$ between 300~K and 100~K;   second, while the gap between the $1xy$ and $2xy$ bands evolves very little with $T$, the spacing between the remaining bands shrinks significantly.  

\begin{figure}[tb]
 \center
  \includegraphics[width=\columnwidth,natwidth=610,natheight=642]{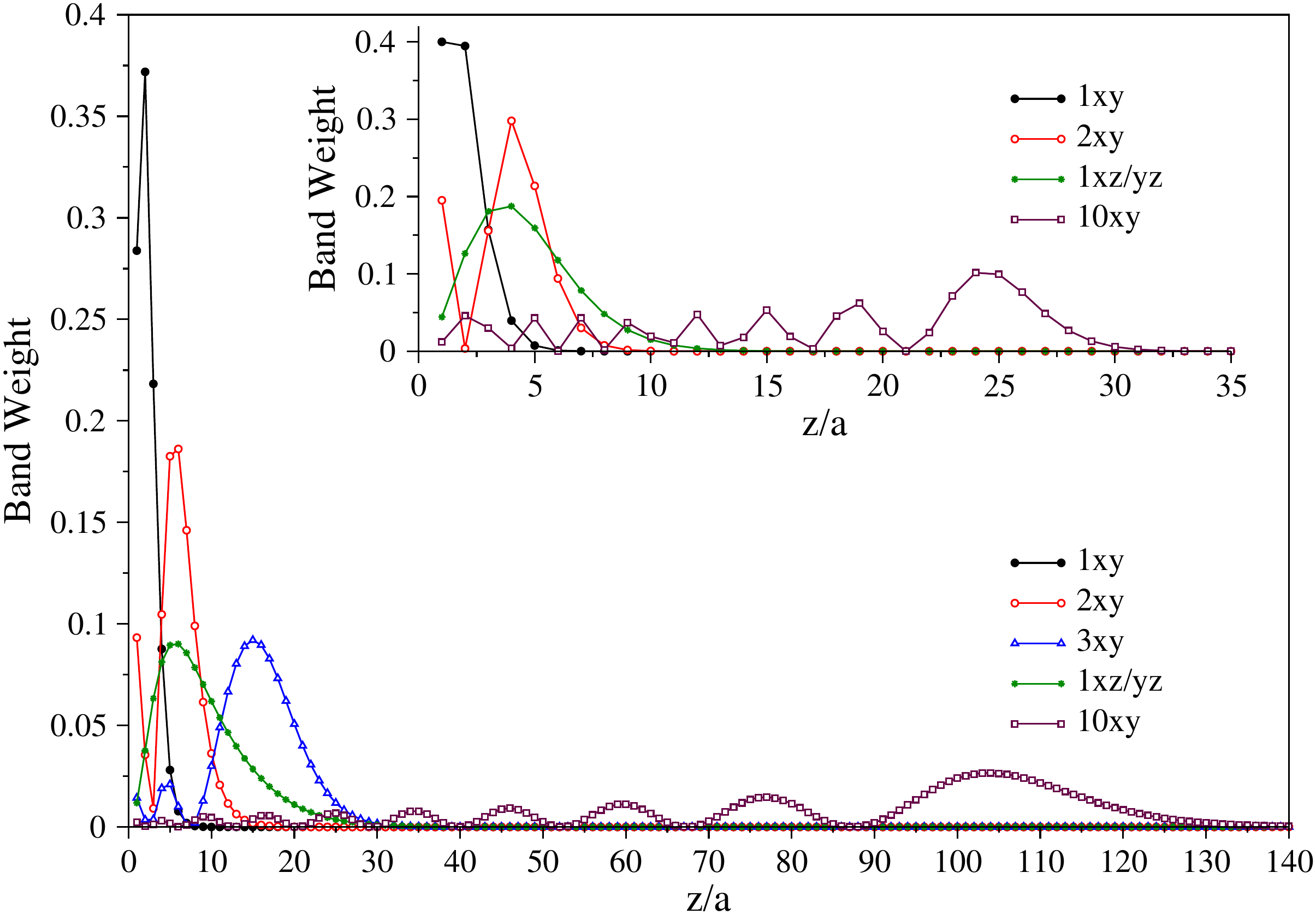}
    \caption{(Color online)  Projected band weights at 10~K (main panel) and 300~K (inset).   The figure shows the band weights of the lowest five bands at 10~K and the lowest four bands at 300~K; these bands contain slightly more than half of the total charge.  For illustration, the band weight of a high-energy $10xy$ band is also shown at each temperature.   The projected weight of band $n$ in layer $j_z = z/a$ for orbital type $\alpha$ is $|\psi_{j_z\alpha,n}|^2$,     where the $\psi_{j_z\alpha, n}$ is the electronic wavefunction. Note that the $xz$ and $yz$ band weights are the same. Results are for $\sigma^s = 0.1e/a^2$.}
   \label{bandwt}
\end{figure}

Coincident with this change in the spectrum, there is a shift of the occupied eigenstates towards three-dimensionality. At 300 K, the  bands shown in Fig.~\ref{band01}(a) have strong 2D character, and the eigenstates are localized within the first 10 STO layers.   This is illustrated in Fig.~\ref{bandwt}, which shows the projected weight $|\psi_{j_z \alpha, n}|^2$ of the first few sub-bands.
Figure~\ref{bandwt} shows that the $1xy$ band is localized within 5 layers of the interface at all temperatures, but that the $2xy$ and $1xz/yz$ bands extend twice as far into the STO at 10~K as at 300~K.  Higher bands are affected even more by temperature, and the $10xy$ band extends four times as far into the STO at 10~K as it does at 300~K.

The distribution of charge amongst the bands is also $T$-dependent.  At 300~K, 57\% of the charge is contained in the first 4 bands ($1xy$, $2xy$, $1xz/yz$); at 10~K, this charge is shared amongst the lowest 5 bands (including $3xy$).  Thus, charge spreads away from the interface as $T$ is lowered for two reasons:  first, occupied bands become less confined; and second, the density of  bands increases, such that higher bands with larger spatial extent become occupied.  

 In particular, the band structure in Fig.~\ref{band01}(d) shows evidence for coexisting 2D and 3D components to the electron gas:   states that are confined to the interface region are characterized by bands that are clearly separated from each other at $\bk = 0$, while 3D states are characterized by a dense continuum of bands.   Indeed, we have found that the first half-dozen bands do not change much with the STO slab thickness $L$, indicative of quantum interface states; however, the sub-band structure at  energies $\gtrsim \mu$ becomes denser as $L$ increases, indicating that these states extend to the back wall of the STO slab, even for $L=200$.   Figure~\ref{band01} thus reinforces the narrative that there is a transfer of electrons from 2D quantum states localized within $\sim 10$ unit cells of the interface to extended 3D tails as $T$ is lowered.

 \begin{figure}[tb]
    \centering
    \includegraphics[width=\columnwidth,natwidth=610,natheight=642]{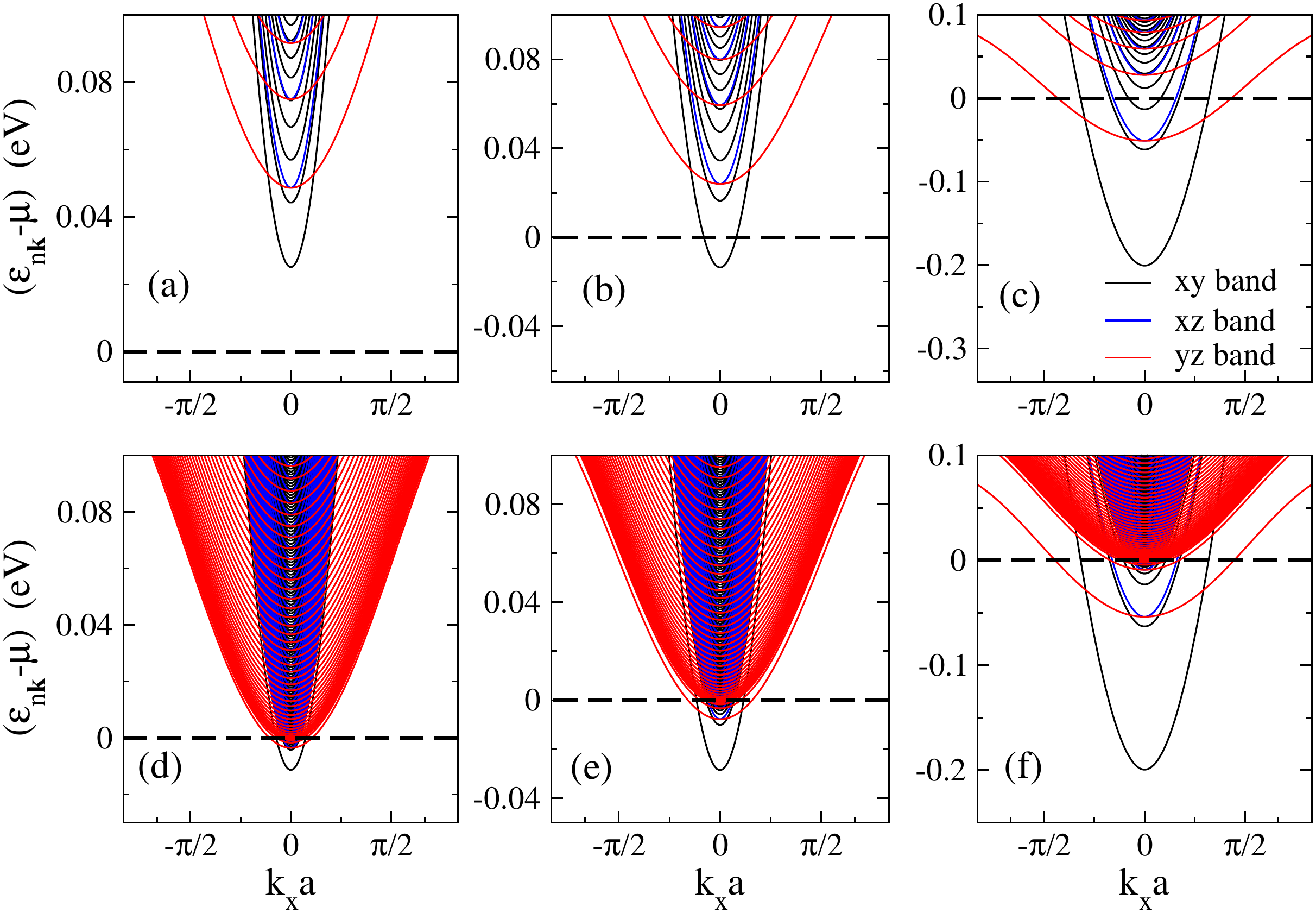}
    \caption{(Color online) Doping- and temperature-dependent band structure of a STO interface. Results are for (a)-(c) 300 K, and (d)-(f) 10 K.  Doping levels are (a), (d) $\sigma^s= 0.05e/a^2$; (b), (e)
    $\sigma^s= 0.1e/a^2$; (c), (f) $\sigma^s= 0.5e/a^2$.}
    \label{banddop}
\end{figure}

Figure~\ref{banddop} compares the calculated band structures at low and high temperature for low, intermediate, and high doping.  At all electron densities, the visible portions of the spectra comprise a set of distinct bands with 2D character at 300~K.  At 10~K, the spectra consist of a small number of low-energy 2D bands that are clearly separated from a 3D continuum with $\epsilon_{n\bk} \gtrsim \mu$.   The low-energy bands are the source of the interfacial component of the charge density in Fig.~\ref{n01}.   Consistent with Fig.~\ref{n01}, the 2D bands at high doping [Fig.~\ref{banddop}(c) and (f)] are nearly independent of $T$.

In summary, we find that there is a discrete spectrum of quantum 2D states that are confined to within 10 unit cells of the interface, and a higher energy continuum of 3D states that extend hundreds of unit cells into the STO.   The principal result of this section is that the 3D states lead to a partial deconfinement of the electrons from the interface at low $T$, and that this deconfinement becomes more pronounced as the total 2D electron density is reduced.

\subsection{Spectral function}
\label{sec:arpes}

The temperature-dependent band structure can be observed by ARPES, and indeed recent ARPES experiments at low temperature have found features consistent with the predicted band structure.\cite{Cancellieri:2013wa,Cancellieri:2016fw}
ARPES is a surface-sensitive technique that measures 
the projection of the spectral function onto the top STO layer; furthermore, 
photon polarization can be used to selectively probe different orbital symmetries.  
For direct comparison  we therefore calculate $A_{i_z,\alpha}(\omega, {\bf k})$, 
the projected spectral function in layer $i_z$ for orbital type $\alpha$.  This is given by
\begin{equation}
 A_{i_z,\alpha}(\omega, {\bf k})=\sum_{n}| \psi_{i_z\alpha, n}({\bf k})|^2 \delta(\omega-\epsilon_{n\bf k}),
 \label{eq:Akw}
\end{equation}
where $ | \psi_{i_z\alpha, n}({\bf k})|^2$ is the weight of the $n$th band in layer $i_z$ for orbital type $\alpha$,  and $\epsilon_{n\bf k}$ is the dispersion of the $n$th band.  
The delta-function has a Lorentzian broadening of 0.01 eV, 
which is comparable to the energy resolution of high-resolution ARPES experiments.

\begin{figure}[tb]
    \centering
    \includegraphics[width=\columnwidth]{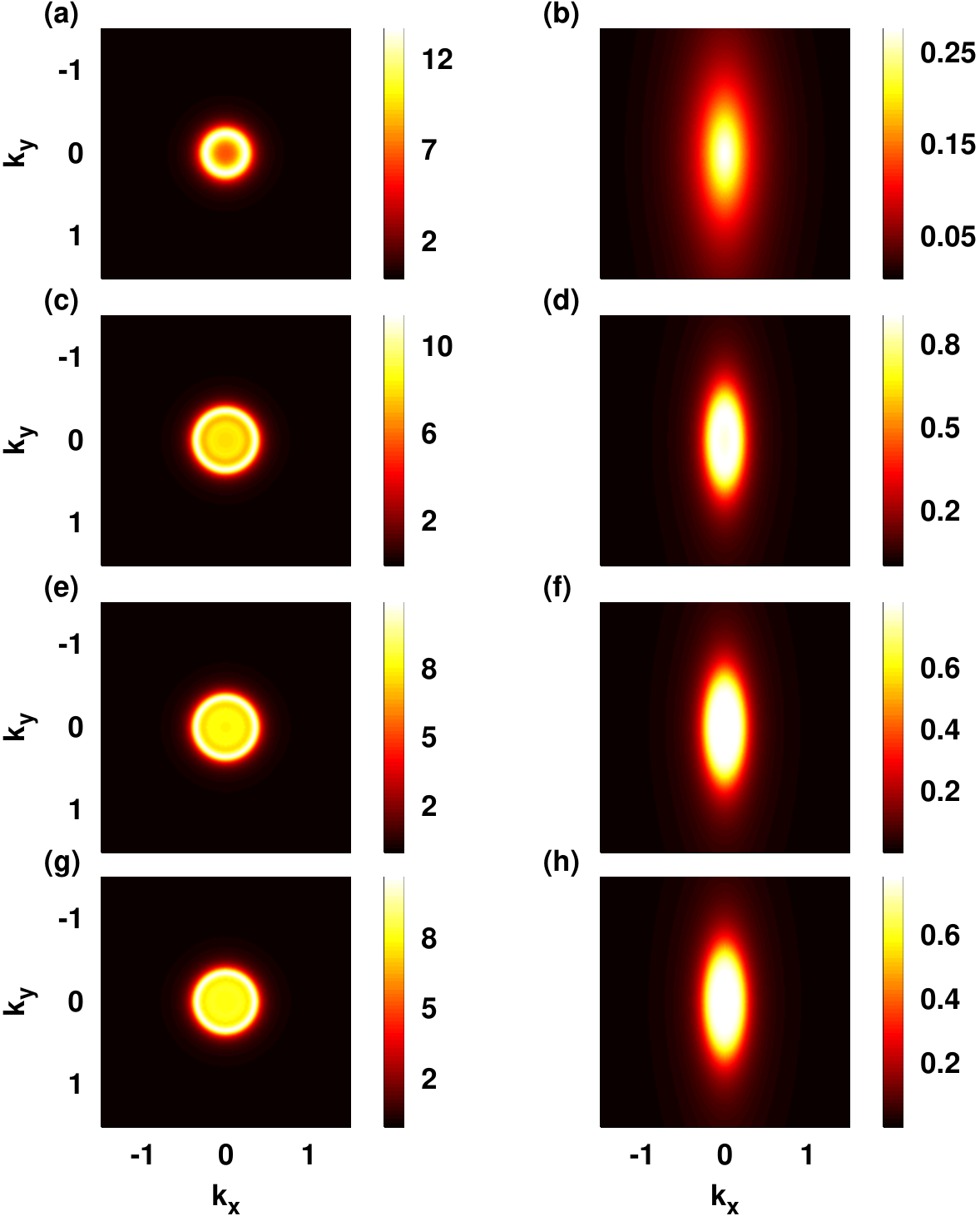}
    \caption{(Color online) Projected  spectral function at the interface for quasiparticle energy $\mu$. The left panels show $A_{1,xy}(\mu,\bk)$ for $xy$ bands at (a) 300~K, (c) 100~K, (e) 50~K, and (g) 10~K. The right panels present the corresponding spectral function $A_{1,xz}(\mu,\bk)$.  Results are for  $\sigma^s=0.1e/a^2$.}
   \label{Spec01}
\end{figure}

We are principally concerned with two main points about the spectral function: 
the intensity of the various features of the band structure, 
which is nominally related to the weight of the different bands at the surface; and 
the size of the apparent Fermi surfaces, which is nominally related to the filling of each band.
Because both the band weight and band dispersion change with temperature, 
as shown in Figs.~\ref{band01} and \ref{bandwt}, we expect that the projected spectral function must also change with temperature.  

We begin with the case of intermediate electron density.
Figure~\ref{Spec01} shows the temperature-dependent spectral function $A_{1,\alpha}(\mu, {\bf k})$ 
at the interface ($i_z=1$) for quasiparticles at the electrochemical potential $\mu$. 
The left panels present the evolution of the projected spectral function for the $xy$ bands;
the right panels show the corresponding spectral function for the $xz$ bands.  (The spectral functions for the $yz$ bands can be obtained by rotating the $xz$  image by $\pi/2$.)

At  300~K, we observe an intense ring with $xy$ symmetry, corresponding to the $1xy$ band [Fig.~\ref{Spec01}(a)], and a very weak cigar-shaped feature associated with the $1xz$ band [Fig.~\ref{Spec01}(b)].  The disparity between the $xy$ and $xz/yz$ intensities is consistent with the fact that only the $1xy$ band crosses $\mu$ at this high temperature.  Indeed, the bottom of the $1xz$ band is $\sim 0.035$ eV above $\mu$, and is only observable in Fig.~\ref{Spec01}(b) because of the finite energy resolution in Eq.~(\ref{eq:Akw}).

At 100~K, the intensity of the $1xy$ band decreases slightly, and an intense disk centered at $\bk=0$ appears [Fig.~\ref{Spec01}(c)]. This change in the spectral function reflects both changes in the band structure and a shift of the chemical potential to higher energies [c.f.\ Fig.~\ref{band01}(b) and (c)].  
At this temperature, multiple $xy$ bands pass within 0.01~eV of the chemical potential; while the $1xy$ band appears as a distinct ring, these remaining $xy$ bands blur together to form a disk.  The $1xz$ band [Fig.~\ref{Spec01}(d)] continues to be an order of magnitude less intense than the $xy$ bands, despite the fact that the $1xz$ band dispersion crosses $\mu$ at 100~K. This is because of the small weight of the $1xz$ band at the interface [Fig.~\ref{bandwt}].

Below 50~K, the intensity of the $1xy$ ring does not change [Fig.~\ref{Spec01}(e) and (g)], but the disk intensity increases slightly because higher energy $xy$ bands shift downwards as $T$ decreases, as shown in Figs. \ref{band01}(c) and (d).   At the lowest temperatures, this disk represents the projection of the 3D tail states onto the surface.  The intensity of the $xz$ bands remains an order of magnitude smaller than that of the $xy$ bands [Fig.~\ref{Spec01}(f)].  There is very little change to the apparent spectrum below 50~K.

Focusing on bands of $xy$ symmetry, we note that the apparent filling as determined from the area of the $1xy$ ring is temperature-dependent, and changes by $\sim 20\%$ between 300~K and 100~K.  This change does not reflect a 20\% change in the filling of the $1xy$ band however, because of the rather large change in $\mu$, which shifts upwards by almost 0.02~eV as $T$ is lowered.  Below 100~K, the ring's surface area does not  significantly change with temperature. 

\begin{figure}[tb]
    \centering
    \includegraphics[width=\columnwidth]{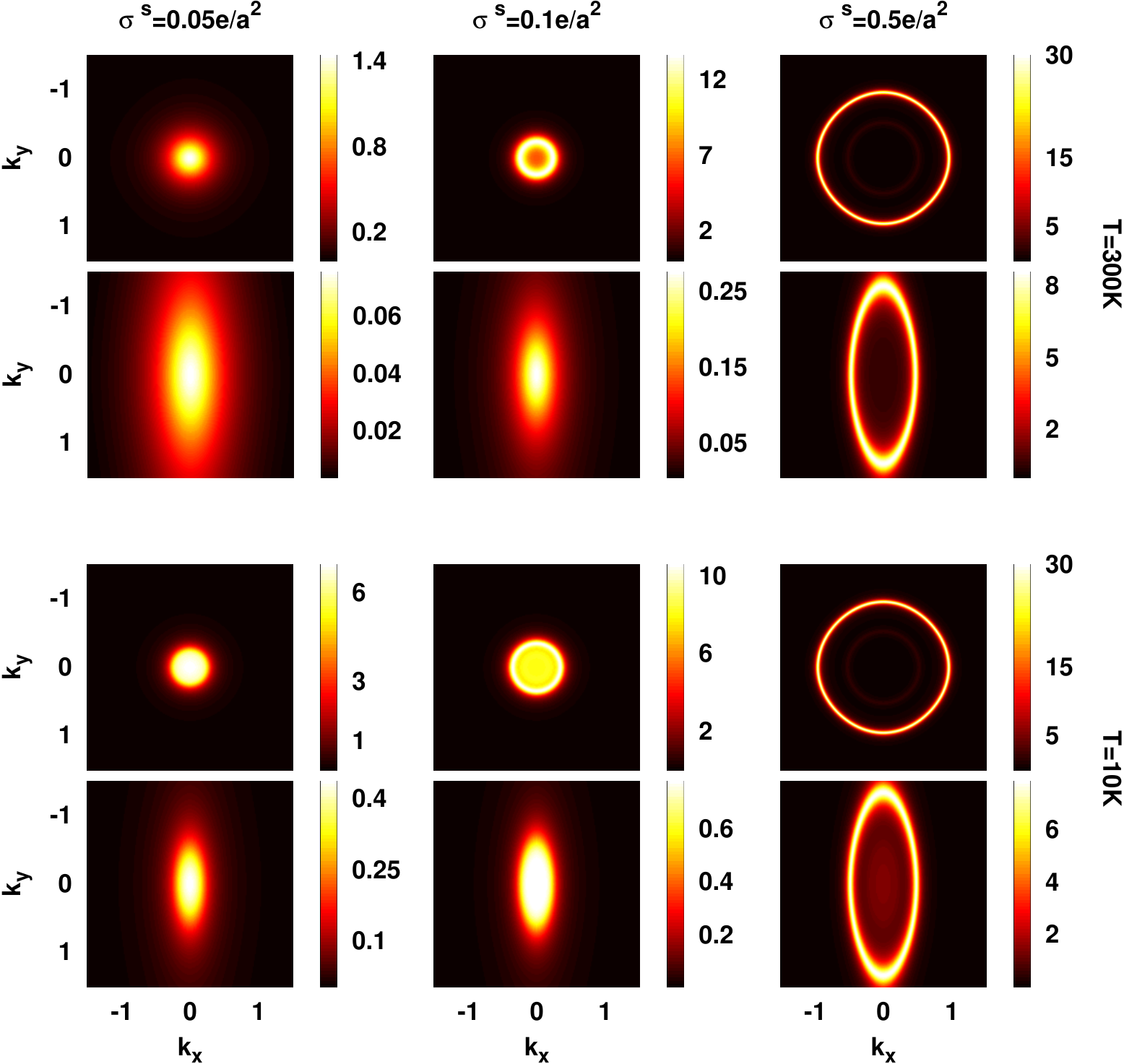}
    \caption{(Color online) Projected spectral function at low, intermediate, and high electron densities, and at $T=10$~K and $T=300$~K.  Results are shown for $xy$ bands (rows 1 and 3) and $xz$ bands (rows 2 and 4) bands. }
   \label{Spec}
\end{figure}

Next, the doping-dependence of the spectral function is shown in Fig.~\ref{Spec}.  As expected,  the surface area of the bands increases with $\sigma^s$, in agreement with  Ref. \onlinecite{Cancellieri:2013wa}; however, it is the temperature-dependence of the intensity that is most striking.  The spectral function is almost independent of $T$ at $\sigma^s = 0.5 e/a^2$, which is a direct result of the strongly nonlinear dielectric response in the interface region at high doping.  In contrast, at low doping, the intensity of the spectral function at $\mu$ is strongly $T$-dependent, primarily because of the strong $T$-dependence of the chemical potential.

Several groups have performed ARPES experiments on STO interfaces at low temperatures, 
and the shapes and surface areas of our calculations are in good agreement with the measured Fermi surfaces for approximately the same doping.\cite{Berner:2013kp,Cancellieri:2013wa,Cancellieri:2016fw}  Notably the $xz$ (and $yz$) bands are more than an order of magnitude weaker than the $xy$ bands in our calculations; and while the relative intensities of the bands observed in ARPES depend on matrix elements,  the $d_{xz/yz}$ bands are indeed considerably weaker than the $d_{xy}$ bands.\cite{Berner:2013kp} 

In summary, our calculations agree with ARPES experiments at low temperatures, and we make two predictions regarding spectral function $A_{1,\alpha}(\mu,\bk)$ at high temperatures:  first that the area of the $1xy$ ring should shrink as $T$ is raised above 100 K;  and second that the intensity of the $d_{xz/yz}$ bands should drop dramatically above 100 K.

\subsection{Local and nonlocal dielectric functions}
\label{sec:local_v_nonlocal}
We finish Sec.~\ref{sec:results} with a brief discussion of the dielectric model used in this work.  The dielectric response obtained from Eq.~(\ref{u}) contains both nonlocal and nonlinear contributions to the polarization.  The nonlinearity has been discussed previously,\cite{Khalsa:2012fu,Reich:2015ut,Peelaers:2015fh} and was generally found to be important only near the  interface for $\sigma^s \gtrsim 10^{14}$ $e$/cm$^2$, consistent with our findings here.  In this section, we investigate the effects of the nonlocal dielectric response on $n(z)$.
 
We compare the charge density profile obtained from the nonlocal  matrix of force constants $D_{i_zj_z}$, defined previously, with the one obtained from a local  matrix ${\tilde { D}}_{i_zj_z}={\tilde { D}}_{i_z} \delta_{i_z,j_z}$.  For purposes of comparison, we choose $\tilde D_{i_z}$ such that it gives the same linear response for a uniform  electric field as $D_{i_zj_z}$.  If the electric field $E_{l_z}$ and normal coordinate $u_{j_z}$ are independent of position in Eq.~(\ref{u}), we obtain in the weak-field limit
\begin{eqnarray}
QE &=& \sum_{j_z} D_{i_z j_z} u \nonumber  \\
&=& D_{k_z=0} u.
\end{eqnarray}
To obtain the same result for $\tilde D_{i_z j_z}$, we define ${\tilde { D}}_{i_z}=D_{k_z=0}$.

\begin{figure}[tb]
    \centering
    \includegraphics[width=\columnwidth,natwidth=610,natheight=642]{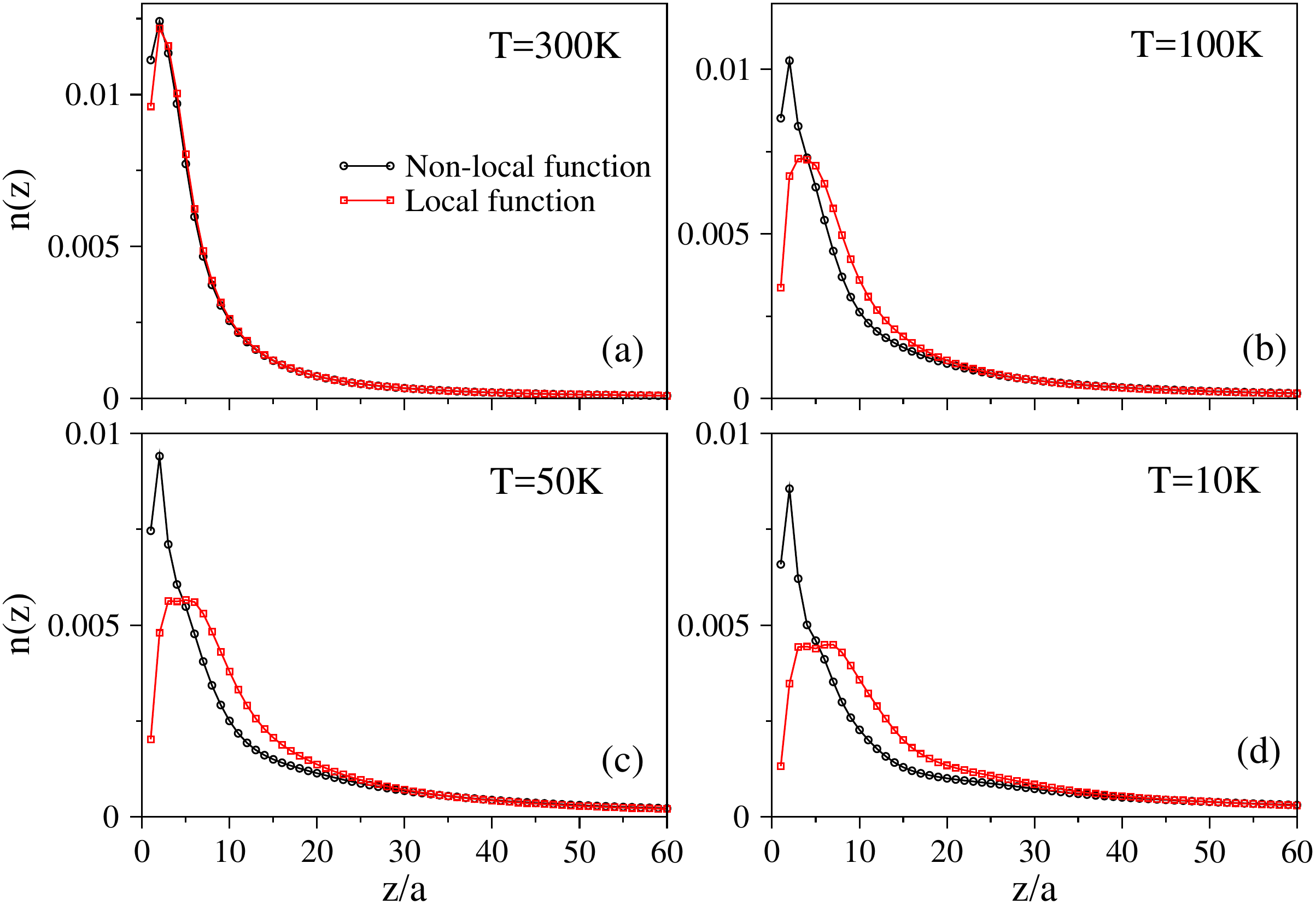}
    \caption{(Color online) Comparison of local and nonlocal models for the dielectric response.  The charge density profile for the two models is shown at  temperatures (a) $T=300$~K, (b) 100~K, (c) 50~K, and (d) 10~K  with $\sigma^s=0.1e/a^2$.  The first 60 layers of an $L=200$ layer thick STO slab are shown.}
    \label{nloc/loc}
\end{figure}

Figure \ref{nloc/loc} shows the charge density profile at different temperatures for local and nonlocal  force constants.   At 300~K,  the two give the same charge density profile [Fig.~\ref{nloc/loc}(a)].  However, as the temperature is lowered, charge moves away from the interface more rapidly for the local case than for the nonlocal case [Fig.~\ref{nloc/loc}(b)-(d)].  Far from the interface,  both cases yield identical results as found in Ref.~\onlinecite{Reich:2015ut};  this is because we defined $\tilde D_{i_z}$ such that it gives same homogeneous response as $D_{i_zj_z}$.

The behavior shown in Fig.~\ref{nloc/loc} can be understood simply.  The dielectric response is connected to a soft optical phonon mode with dispersion $\omega_\bk$ satisfying $D_\bk = M\omega_\bk^2$ where $M$ is the effective mass of the mode.  At high temperatures, $\omega_\bk$ has a relatively smooth dispersion; however the dispersion, and consequently $D_\bk$, develops a sharp feature at low $T$ as the mode softens near $\bk=0$.\cite{Cowley:tr}    From the properties of Fourier transforms, it follows that the range of $D_{i_zj_z}$ is therefore greater at low $T$ than at high $T$, or equivalently that the response is more local at high $T$.  This accounts for the similarity between the two models at 300~K.  The different charge profiles that emerge at low $T$ indicate that the local dielectric function is more effective at screening the electric field in regions where there are strong field gradients.  

\section{Discussion and Conclusions}
\label{sec:discussion}
The calculations in this work are based on a combination of two established models:  the dielectric properties of the STO are modeled by a Landau-Devonshire free energy similar to those used to describe the insulating parent compound,\cite{Dec:2005cr,Palova:2009js} while the electronic properties are described by a tight binding model, similar to what is done elsewhere.\cite{Stengel:2011hy,Zhong:2013cr} Unlike conventional semiconductors, the STO dielectric function is strongly temperature- and electric field-dependent. This leads to counterintuitive behavior at STO interfaces; namely, that the electron gas is more strongly confined at high temperatures and electron densities than at low temperatures and electron densities. Consequently,  our calculations make predictions that differ from commonly held views regarding the electron distribution in STO interfaces.

The conventional view is that the electronic properties are dominated by quantum 2D states, and indeed experiments find that the majority of the charge is bound to within $~\sim 10$ nm of the interface.\cite{Reyren:2009va,Copie:2009ev,Basletic:2008ja,Dubroka:2010bi}   Measurements of the nonlinear Hall coefficient have been modeled by two occupied sub-bands: a low-mobility band containing most of the conduction electrons, and a high-mobility band containing a minority of carriers.  The mobilities of the two components vary from sample to sample, and may differ by orders of magnitude.\cite{Joshua:2012bl,Lerer:2011bp,Jost:2014uz,Kim:2010fl,Guduru:2013iz}  While the two-band interpretation is conceptually useful, it has been noted that inconsistencies within the two-band analysis suggest a more complicated band structure.\cite{Joshua:2012bl}   At low electron densities, the picture is clearer: experiments have found a Lifshitz transition near electron densities of $1.5\times 10^{13}$ cm$^{-2}$,\cite{Joshua:2012bl} which is slightly above the metal-insulator transition at $\approx 10^{13}$ cm$^{-2}$.  Below the Lifshitz transition, the magnetic field-dependence of the Hall resistivity is linear, indicating that only a single band is occupied.  

In contrast, the results reported in this work find a large number of occupied bands at all doping levels, similar to previous calculations.\cite{Copie:2009ev,Stengel:2011hy,Khalsa:2012fu,Park:2013gf}  A significant fraction of the occupied bands corresponds to the quasi-3D tail states that extend hundreds of unit cells into the STO substrate.  While the fraction of charge contained in the tails is small at high electron densities, it is over 50\% at low electron densities (Fig.~\ref{n01}).  Perhaps more interestingly,  we have found a strong temperature dependence to the charge distribution at intermediate electron densities, with a pronounced shift of charge into the tails as $T$ is lowered.  The general trend that the charge spreads out as $T$ decreases was observed experimentally;\cite{Copie:2009ev} however, experimental confirmation of quasi-3D tails remains lacking. Indeed, direct observation of the tails may be difficult because, except at the lowest doping levels, the electron density $n(z)$ in the tails is at least an order of magnitude smaller than in the 2D component of the electron gas (Fig.~\ref{n01}).

The tails may be most relevant to transport experiments, since interfacial disorder (eg.\ cation intermixing) is thought to severely reduce the mobility of 2D states near the interface.  A proper comparison between theory and experiment requires a detailed disorder model, which is beyond the scope of this work.  Nonetheless, we can make a few simple observations based on a crude model for the mobility $\mu_{n}$ of the first few bands ($n= 1xy$, $1xz/yz$, $2xy$).  This model assumes that interfacial disorder (eg.\ cation intermixing) is the dominant scattering mechanism and that interband scattering can be neglected.   These assumptions break down at low doping, first because the interband spacing becomes less than the scattering rate, and second because low-lying bands become part of the 3D continuum and are therefore subject to scattering by defects in the STO substrate.  The model is also limited because it provides no information about the mobility of the 3D tails.  For qualitative purposes, however, we can assume that the tails behave similarly to bulk STO.

\begin{figure}
\includegraphics[width=\columnwidth]{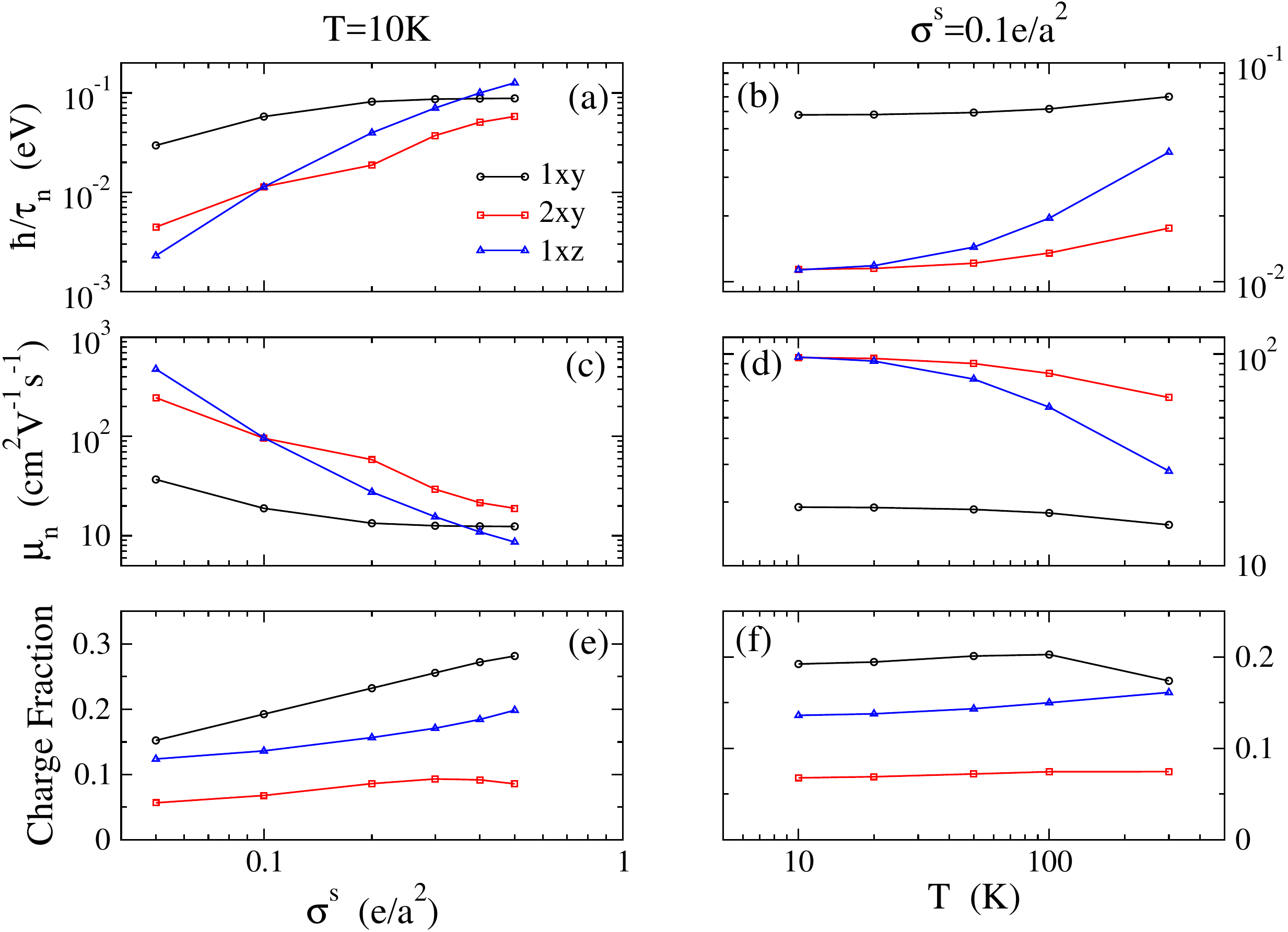}
\caption{(Color online) Transport properties of 2D interface states as a function of (a), (c), (e) 2D charge density at fixed temperature, and (b), (d), (f)  temperature at fixed 2D charge density.    (a), (b) Scattering rate $\hbar/\tau_n$; (c), (d) mobility $\mu_n$; and (e), (f)  fraction of the total charge in band $n$ for $n=1xy$, $2xy$, and $1xz$.  The calculations assume that elastic scattering comes predominantly from  interfacial disorder (eg.\ cation intermixing), and that interband scattering process can be neglected.  Contributions from the 3D tails are not included in this figure.
}
\label{fig:cmpr_to_expt}
\end{figure}

 The simplest ansatz  is to take a quenched disorder model in which the Ti site potentials in the first $\lambda$ STO layers adjacent to the interface are chosen from a random box-distribution of width $W$.  Experimentally, cation intermixing is found to extend over a few unit cells,\cite{Nakagawa:2006gt} and for concreteness, we arbitrarily take $W=1$~eV and $\lambda=2$; however, the qualitative results do not depend strongly on this choice.  Within a Born approximation the electron lifetime $\tau_{n}$ in band $n$ is 
\begin{equation}
\frac{\hbar}{\tau_{n}} = \frac {\sqrt{m_{x,n}m_{y,n}} W^2 a^2}{24\hbar^2 } \sum_{i_z=1}^\lambda |\Psi_{i_z \alpha,n}|^2,
\label{eq:taun}
\end{equation}
where $m_{x,n}$ and $m_{y,n}$ are  effective mass components for  band $n$. The mobility for transport in the $x$-direction is $\mu_n = e \tau_n/m_{x,n}$.  The absolute values of the mobility, which depend on our arbitrary choice of  $W$, are not especially meaningful; however, the trends with doping and temperature shown in Fig.~\ref{fig:cmpr_to_expt} are.  
 
Equation~(\ref{eq:taun}) shows that individual bands' scattering rates depend on the projected band weight $ |\Psi_{i_z \alpha,n}|^2$ onto layers adjacent to the interface.  Two clear trends in Fig.~\ref{fig:cmpr_to_expt}, namely that $\mu_n$ increases when either $\sigma^s$ or $T$ is reduced,  can be traced back to  shifts of the band weight away from the interface (recall, for example, Fig.~\ref{bandwt}).   Similarly, Fig.~\ref{fig:cmpr_to_expt} shows that at fixed $T$ and $\sigma^s$ the mobilities of different 2D bands may differ by an order of magnitude or more because of they have different band weights at the interface.  

While significant, the differences in mobilities between bands that are shown in Fig.~\ref{fig:cmpr_to_expt} are much less than the three orders of magnitude difference between high- and low-mobility electrons reported in Refs.~\onlinecite{Guduru:2013iz,Jost:2014uz}.  Those experiments instead suggest that the two electronic components live in different environments.  With this in mind, we speculate that the low-density high-mobility component of the electron gas observed over a wide range of electron dopings,\cite{Joshua:2012bl,Lerer:2011bp,Jost:2014uz,Kim:2010fl,Guduru:2013iz}  may in fact correspond to the 3D tails in our calculations.  These tails have very little overlap with the interface, and the scattering of conduction electrons will be determined by the defect density in the STO substrate.   The remaining high-density low-mobility component of the electron gas then must correspond to the 2D interface states, whose mobility is limited by interfacial disorder.   We point to three experimental observations that are broadly consistent with this proposed scenario:  
\begin{itemize}
\item First, our calculated charge densities in the interface and tail regions roughly correspond to the observed fractions of low and high mobility charges.  Ref.~\onlinecite{Guduru:2013iz} reports that for high electron densities, the high-mobility component of their electron gas comprises less than 10\% of the total electron density, while Ref.~\onlinecite{Lerer:2011bp} found that at intermediate densities  the high-mobility component contains a third of the total electron density.  Similarly,  Fig.~\ref{n01} shows that the fraction of the total charge in the tail region at 10~K rises from less than 10\% at high electron density to roughly 50\% at intermediate density.  
\item Second, the predicted temperature dependence of the mobility is qualitatively consistent with available experiments.  At intermediate electron densities, Ref.~\onlinecite{Lerer:2011bp} found that the conductivity of the high-density component is nearly independent of $T$ (up to 30~K), while the conductivity of the low-density component drops by an order of magnitude.  Similarly, Fig.~\ref{fig:cmpr_to_expt} shows that the mobilities of the interface states are almost constant between 10~K and 30~K, owing to modest changes in the confinement of their wavefunctions to the interface.  Conversely, we expect the tail states to exhibit a strong temperature-dependence, assuming that they follow the behavior of bulk STO.\cite{Spinelli:2010dm,Faridi:2016ue}   
\item Third, at low electron densities, Ref.~\onlinecite{Joshua:2012bl} argued that the electrochemical potential is pinned to the bottom of a heavy band that acts as a charge reservoir.  They speculated that this reservoir consists of interfacial $d_{xz/yz}$ bands; however, our calculations find that at 10~K the electrochemical potential is pinned to the bottom of the quasi-3D tail bands (Fig.~\ref{banddop}).  Because the density of states in the tails is extremely high compared to the 2D interface states, we argue that the tails provide a more natural explanation for the observed charge reservoir.
\end{itemize}

We note that there are open questions that are not addressed by the simple arguments presented here.  Our model does not predict the Lifshitz transition observed by Ref.~\onlinecite{Joshua:2012bl} at low electron density, for example.  Instead, the $1xy$ band in our calculations continuously merges with the 3D continuum as the electron density is lowered.  We do not know the reason for this discrepancy, although our neglect of spin-orbit coupling, which is known to be important at low doping, is an obvious candidate.  It is also not yet clear whether the multiple occupied bands predicted by our calculations are consistent with the two-band interpretation of transport coefficients; in particular, a proper calculation of magnetoresistance with a qualitatively accurate disorder model is required to understand the extend to which our model is compatible with experiments.

Finally, we remark that our calculations have implications for the superconducting state that has been observed at STO interfaces.  This state has been shown to be 2D, with a characteristic thickness of $\sim 10$~nm inferred from measurements of the critical magnetic field anisotropy.\cite{Reyren:2009va}  While this naively seems to contradict the prediction of quasi-3D tails that extend hundreds of unit cells into the bulk, we note that bulk STO is superconducting for 3D electron densities between $6\times 10^{-4}$ and $2\times 10^{-2}$ electrons per unit cell,\cite{Edge:2015fj}  such that the lowest density regions of the tail region are not expected to be superconducting.  For our ``typical'' case of $\sigma^s = 0.1 e/a^2$, Fig.~\ref{n01}(b) suggests that superconductivity extends roughly 30 unit cells into the STO substrate, in agreement with experiments.

In summary, we have explored the temperature- and doping-dependent band structure of model STO interfaces.  The calculations presented in this work suggest a significant role for quasi-3D tail states, contrary to a widely held perception that the interfaces are dominated by 2D states.  These tail states extend hundreds of unit cells into the STO substrate, and are extremely sensistive to both electron doping and temperature.  We have shown that photoemission experiments can be used to probe the temperature-dependent band structure; however, the tail states exist far from the interface and are therefore invisible to ARPES.  We speculate, however, that the tail states are key to understanding transport experiments, and have provided some qualitative evidence to support this idea.

\section*{Acknowledgments}
We acknowledge support by the Natural Sciences and Engineering Research
Council (NSERC) of Canada.

\appendix
\section{Fitting the dielectric model to experiments}
\label{app:A}
In this appendix, we outline the process by which the model parameters were fitted to experimental measurements of the field- and temperature-dependent dielectric susceptibility
\begin{equation}
\chi_{ij}(T,E) =\frac{1}{\epsilon_0}\frac{\partial P_i}{\partial E_j},
\end{equation}
where $i$ and $j$ label unit cells.
For a uniform electric field, the polarization and normal coordinate $u$ are also uniform, and
from Eq.~(\ref{eq:Pu}),
\begin{equation}
\chi(T,E) =\frac{Q}{\epsilon_0 a^3}\frac{\partial u}{\partial E}
\end{equation}
 From Eq.~(\ref{u}), we then obtain
\begin{equation}
D_{\bq=0} u+\gamma u^3=QE,
\label{eq:u}
\end{equation}
where $D_{\bq=0} = \sum_j D_{ij}$.  
Differentiating Eq.~(\ref{eq:u}) with respect to $E$, we obtain
\begin{equation}
\chi(T,E) =\frac{Q^2}{\epsilon_0 a^3}\frac{1}{D_{\bq=0}+3\gamma u^2},
\label{eps}
\end{equation}
where $u$ is obtained from Eq.~(\ref{eq:u}).

\begin{figure}[tb]
 \includegraphics[width=\columnwidth]{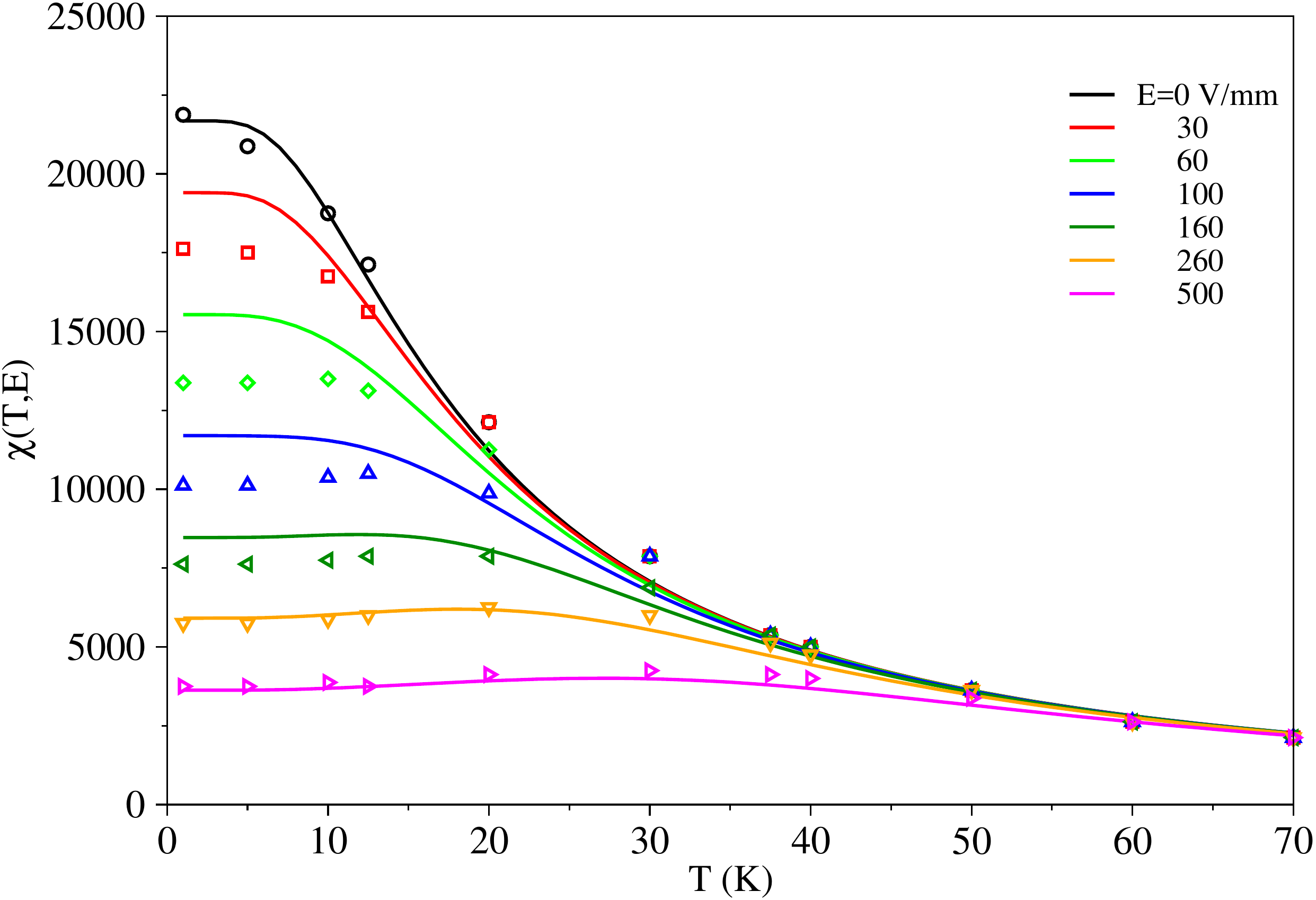}
\caption{(Color online) Comparison of the uniform dielectric susceptibility $\chi(T,E)$ to the experimental results of Dec {\em et al.}\cite{Dec:1999wh}  Symbols are experimental data, solid lines are theory.  Results are shown as a function of temperature for different electric field strengths.}
\label{fig:dec}
\end{figure}

Dec {\it et al.}\cite{Dec:1999wh} showed that the zero-field dielectric susceptibility can be fitted empirically by 
\begin{equation}
\chi(T,0) =\left( \frac{T_0}{T_Q} \right )^\xi
\label{eq:curie}
\end{equation}
where $T_0$ is a constant and ${T_Q=T_s \coth(\frac{T_s}{T})}$ is the quantum analogue of the temperature: when ${T\gg T_s}$, ${T_Q\rightarrow T}$ and when ${T\ll T_s}$, ${T_Q\rightarrow T_s}$. We note that  ${\xi }$ was found to be 2 at low temperatures and ${1}$ at high temperatures; to reduce the number of fitting parameters, we take $1< \xi <2$ to be constant over the entire temperature range.  This improves the quantitative fit to the data, but means that we do not reproduce the correct critical exponents at low $T$.

Equating Eq.~(\ref{eq:curie}) to Eq.~(\ref{eps}) in the zero-field limit yields
\begin{equation}
D_{\bq=0}=\frac{Q^2}{\epsilon_0 a^3}\left( \frac{T_Q}{T_0}\right )^\xi.
\label{eq:D}
\end{equation}
Reinserting this into Eq.~(\ref{eps}) gives us an equation for the nonlinear susceptibility at finite fields with the fitting parameters $T_s$, $T_0$, $Q$, $\xi$, and $\gamma$.  We fit this expression to the experimental data of Ref.~\onlinecite{Dec:1999wh}, and the result is shown in Fig.~\ref{fig:dec}. The model reproduces the data at both low  and room temperatures and a range of electric fields from ${0 ~\textnormal{V/mm}}$ to ${500 ~\textnormal{V/mm}}$ with a maximum relative error of ${16\%}$.  The best fit parameters  are given in Table~\ref{cons}.



To extend this model to finite ${\bq}$, we take the empirical expression\cite{Khalsa:2012fu}
\begin{equation}
D_\bq=M [\omega_0^2-\omega_1^2e^{\frac{-(\alpha_1 \bq)^2}{2}}-\omega_2^2(T)e^{\frac{-(\alpha_2 \bq)^2}{2}}],
\label{eq:khalsad}
\end{equation}
where $M$ is the reduced mass for the soft mode, $\omega_0$, $\omega_1$, and $\alpha_1$ are used to reproduce the measured phonon dispersion\cite{Cowley:tr} at 90~K, and $\omega_2(T)$ and $\alpha_2$ are used to capture the low-temperature phonon dispersion. Equation~(\ref{eq:w2T}) for the temperature-dependence of $\omega_2(T)$ can be obtained by setting ${\bq=0}$ in Eq.~(\ref{eq:khalsad}) and equating it to Eq.~(\ref{eq:D}).

\section{Finite-size effects}
\label{sec:fs}

\begin{figure}[tb]
    \includegraphics[width=\columnwidth]{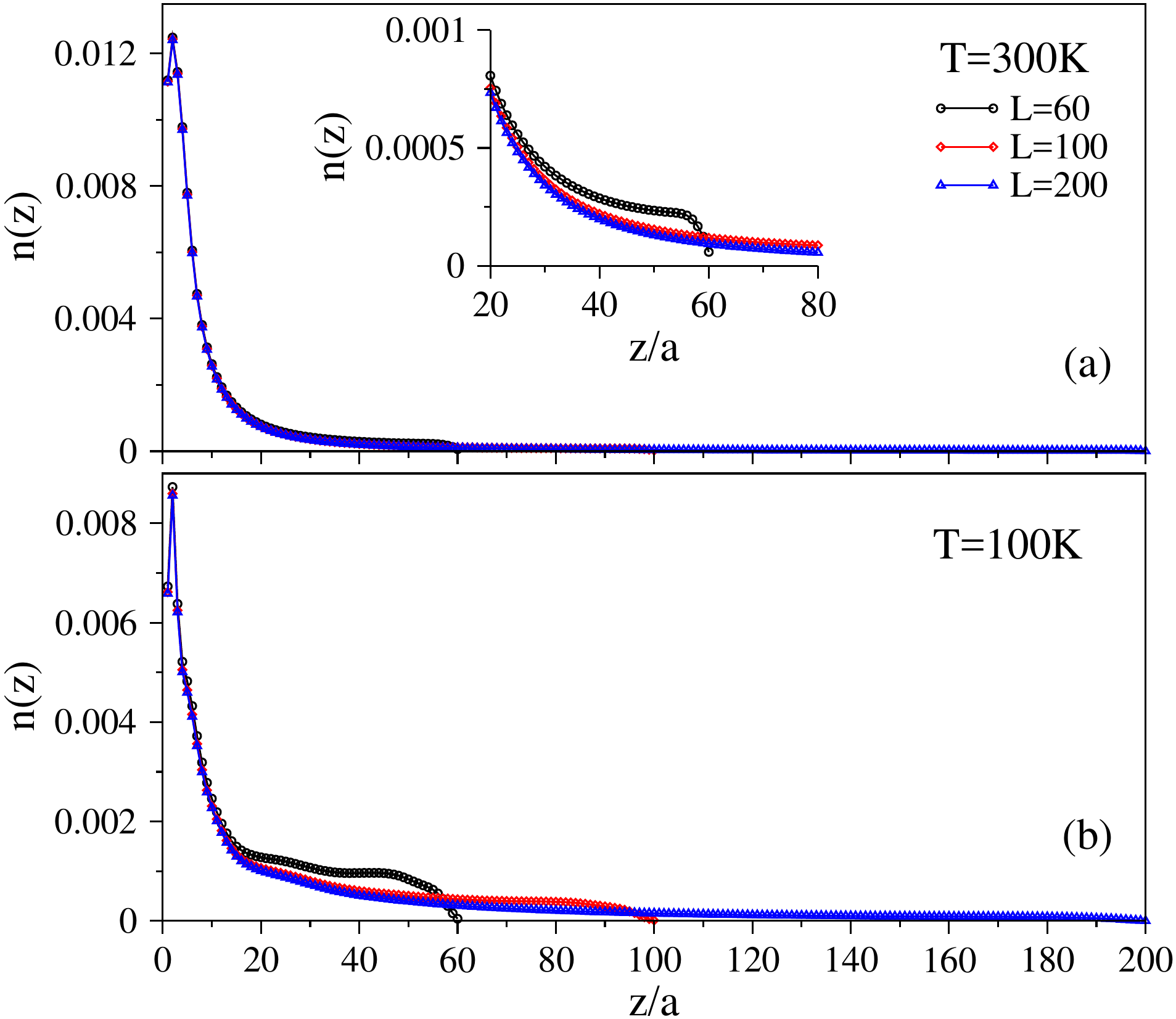}
    \caption{(Color online) Finite-size effects on the electron density $n(z)$.  Results are shown for (a) 300~K and (b) 10~K.    The electron density is given in units of electrons per unit cell.  Results are shown for STO thicknesses of $L=60$, $100$, and $200$ layers.  The 2D charge density is $\sigma^s=6.5\times 10^{13}$ $e$/cm$^2$ (corresponing to $0.1e/a^2$). 
 {\em Inset}. A zoom-in of the charge profile at $300$K is shown. }
   \label{n60-200}
\end{figure}

Most previous numerical simulations (including DFT and tight-binding models) have been restricted to a few tens of STO layers, and it is unclear to what extent they are affected by the thickness of the STO slab. Figure~\ref{n60-200} compares the  electron density $n(z)$ inside the STO slab for different slab thicknesses  ($L=60$, $100$, and $200$ layers) at high and low temperatures.  
For qualitative purposes, we can divide the charge profile to two regions:  one is close to the interface ($z \lesssim 10a$) and has most of total charge, 
and other ($z \gtrsim 10a$) contains a long tail that extends deeply into the STO slab.  Near to the interface, the distribution of charges is nearly identical for all thicknesses at 300 K [Fig. \ref{n60-200}(a)], and depends only weakly on thickness at 10 K [Fig.~\ref{n60-200}(b)].  In contrast, the shape of the long tail changes with the system size, particularly at low $T$.  We note, however, that the total amount of charge in the tail region is roughly independent of slab thickness.  

Density functional theory (DFT) calculations on systems of up to 30 STO layers\cite{Son:2009wb} have shown that the charge profile near the interface decays exponentially with $z$, and that the tails decay algebraically as $n(z) \sim (z-z_0)^{-1}$.  Figure~\ref{n60-200} suggests that the exponential behavior near the interface is robust, but that the tail  is subject to significant finite-size effects; indeed, we find that power law fits to the tails give different results for different system sizes.
 
\bibliography{interfaces}

\end{document}